\documentclass[a4paper, 12pt]{article}
\usepackage{}
\usepackage{amsfonts}
\usepackage{dsfont}
\usepackage{mathrsfs}
\usepackage{amssymb}
\usepackage{bbm}
\usepackage{epsfig}
\usepackage{amsmath}
\usepackage{appendix}
\usepackage{color}
\usepackage[english]{babel}

\usepackage{amsfonts}
\usepackage{dsfont}
\usepackage{mathrsfs}
\usepackage{amssymb}
\usepackage{bbm}
\usepackage{epsfig}
\usepackage{amsmath}
\usepackage{appendix}
\usepackage{float}
\usepackage[colorlinks,linkcolor=blue]{hyperref}
\usepackage{fancyhdr}
\newtheorem{them}{Theorem}[section]
\newtheorem{defn}{Definition}[section]

\newtheorem{lem}{Lemma}[section]

\newtheorem{rem}{Remark}[section]

\title{\bf  Distributed Continuous-time Approximate Projection Protocols for Shortest Distance Optimization
 Problems\footnote{This work was supported by the
National Natural Science Foundation of China under Grant 71401163 and 61333001, Beijing
Natural Science Foundation 4152057,
Hong Kong Research Grants Council under Grant 419511 and Hong Kong Scholars Program under Grant XJ2015049.}}
\date{}

\author{Youcheng Lou,
     Yiguang Hong, Shouyang Wang}

\begin{document}

\maketitle

\begin{abstract}
In this paper, we investigate the distributed shortest distance optimization problem for a multi-agent network to cooperatively minimize
the sum of the quadratic distances from some convex sets, where each set is only associated with one agent.
To deal with the optimization problem with projection uncertainties, we
propose a distributed continuous-time dynamical protocol based on a new concept of approximate projection. Here
each agent can only obtain an approximate projection point on the boundary of its convex set, and communicate with its neighbors over a
time-varying communication graph. First, we show that no matter how large the approximate angle is, the system states are always bounded
for any initial condition, and uniformly bounded with respect to all initial conditions if
the inferior limit of the stepsize is greater than zero. Then, in the two cases, nonempty intersection and empty
intersection of convex sets, we provide stepsize and approximate angle conditions to ensure the optimal convergence, respectively.
Moreover, we give some characterizations about the optimal solutions for the empty intersection case and
also present the convergence error between agents' estimates and the
optimal point in the case of constant stepsizes and approximate angles.
\end{abstract}


\section{Introduction}

In recent years, distributed optimization of a sum of convex functions has attracted much attention
due to its wide applications in resource allocation, source localization,
and robust estimation (referring to \cite{ber,nedic1,nedic2,nedic3,joh09,lu1,lu2,jak}).
A whole optimization task can be accomplished cooperatively by a group of autonomous
agents via simple local information exchange and distributed protocol design
even when the communication graph among agents is time-varying.

Although many existing distributed optimization works have been done by discrete-time algorithms, more and more attention has been paid to continuous-time algorithms in recent years \cite{shi2,wang2,wang,cor14,cor,kva,dro}, partially because the continuous-time models can be
studied by various well-developed continuous-time methods or make the algorithms easily implemented in physical systems.
A distributed continuous-time computation model was proposed to solve an optimization problem
for a fixed undirected graph in \cite{wang2}, with the optimization achieved by controlling the sum of subgradients of
convex functions to make the state enter the optimal solution set, and later this model was generalized to the weight balanced graph case in \cite{cor14},
for differentiable objective functions
with globally Lipschitz continuous gradient. Another continuous-time distributed algorithm with constant stepsize was developed in \cite{kva} for optimization
problems with positivity constraints in a fixed
undirected graph case, where a lower bound of convergence rate and
an upper bound on the agents' estimate error were presented.
Moreover, the relationship between the existing dual decomposition
and consensus-based methods for distributed optimization was revealed in \cite{dro}, where
both approaches were based on the subgradient method, but
one with a proportional control term and the other with an integral control term.

When the optimal solution sets of agents' individual objective functions have a nonempty intersection,
the distributed optimization problem is equivalent to the convex intersection problems (CIP) 
\cite{inter3,inter2,lou13,lou12,shi1,shi2,nedic2}.
A projected consensus algorithm was proposed in \cite{nedic2}
for a network to solve the CIP, and the authors showed that all agents
converge to a common point in the intersection set for weight-balanced and jointly connected
communication graphs.
Later, a continuous-time dynamical system was designed and connectivity conditions were discussed for the optimal convergence in \cite{shi2}.
In addition, a random sleep algorithm was proposed with providing conditions to converge almost surely to a common point in the intersection set in \cite{lou13}, where agents randomly take the neighbor-based average or projection onto their individual sets based on a Bernoulli process.
Almost all the existing optimization results were obtained based on the assumption that the exact projection point onto the convex sets can be obtained
\cite{inter3,inter2,shi1,shi2,nedic2,lin,meng}.

On the other hand, the intersection of the considered convex sets may be empty in practice.   In this case, how to seek a point with the shortest
(quadratic) distance to these sets is also important.  For instance, the supply center location problem is concerned with how to seek the
location of raw materials supply center so that the average transportation cost from the supply center to
the multiple factories is minimal (\cite{handbook,Fra}); the source localization in a sensor network is related to
estimate the location of the source emitting a signal based on the received
signals of multiple sensors in a noisy environment (\cite{zhang15,meng}).  In fact, the problem for both the empty and
nonempty intersection case is referred to as the shortest distance optimization problem (SDOP).
Obviously, CIP is a special case of SDOP, and the average consensus problem is also a special case of SDOP since the optimal solution of the minimum
of the sum of quadratic functions from some points is exactly the average of these points.
Some distributed algorithms were proposed to discuss SDOP.   For example, \cite{meng} formulated the source localization problem as
the SDOP in a plane and proposed a discrete-time distributed algorithm, with
the adjacency matrices of communication graphs required to be doubly stochastic.  Moreover,
\cite{lin} proposed two distributed continuous-time algorithms to solve SDOP in the empty intersection case for connected graphs:
the first one was designed for optimal consensus based on sign functions, and the second one was modified to avoid chattering but only to achieve the optimal consensus approximately.

The objective of this paper is to design a continuous-time distributed protocol to solve SDOP based on approximate projection.
Note that the exact projection is usually
hard to obtain in practice.   Therefore, approximate projection may have to be discussed in different situations, and, in fact, \cite{lou12} proposed a discrete-time approximate projected consensus algorithm to solve
CIP.   The motivation of the current research aims at analysis and distributed design to cooperatively solve SDOP with projection uncertainties and continuous-time dynamics.   For example, in a practical robotic network to solve the SDOP, a continuous-time robot may not always obtain the exact projection point of its own convex set, but only spot some point on the set surface near the exact projection
point.   The contribution of this paper can be summarized as follows.
\begin{itemize}
\item We propose a new concept of approximate projection, which
is related to some points on the convex set's boundary surface and
close to the exact projection point when
the exact projection is hard to obtain.  In other words, we consider an approximate projection related to set boundary surfaces, different from that defined in a ``triangle" in \cite{lou12}. To overcome the analysis difficulties resulting from this new approximate projection, we employ a geometric method to convert the original problem to a heterogeneous stepsize problem.

\item Given any approximate angle, we show that, with the proposed continuous-time algorithm, the agent states
are always bounded for any initial condition, and uniformly bounded with respect to all initial conditions
if the stepsize is not too small. The result with respect to the continuous-time algorithm is different from some results based on some discrete-time ones.
In fact, $\pi$/4 was shown to be the critical approximate
angle for the boundedness of the discrete-time
algorithm with the approximate projection defined in a triangle in \cite{lou12}.

\item  We study SDOP in both the nonempty and empty intersection cases, and propose a unified protocol based on the approximate projection.   In fact, the proposed convergence conditions and proofs in the two cases are quite different.
Note that our result is different from that in \cite{lin} because we handle approximate projections without assuming that the communication graph is always connected, and ours tackles both the nonempty and empty intersection cases, while \cite{lou12} only does the nonempty intersection case.
Moreover, we also discuss the convergence error between agents' estimates and the optimal point
in the case of constant stepsizes and approximate angles. Our results are certainly consistent with those discrete-time algorithms in the literature such as
\cite{nedic1, nedic2} based on the exact projection.
\end{itemize}

The paper is organized as follows. Section 2 shows some basic
concepts and preliminary results.
Section 3 defines an approximate projection concept and formulates our shortest distance optimization problem (SDOP), followed by
Section 4 for the discussions on boundedness and stepsizes.
Section 5 presents the main convergence results for the nonempty intersection case,
while Section 6 for the empty intersection case.
Section 7 discusses the constant stepsize and approximate angle case.
Then Section 8 provides numerical simulations.
Finally, Section 9 gives some concluding remarks.

\emph{Notations:}
$\otimes$ denotes the Kronecker product;
$\textbf{1}$ denotes the vector with all ones;
$(A)_{ij}$ denotes the $i$-th row and $j$-th column entry of matrix $A$;
$y^T$ denotes the transpose of a vector $y\in \mathbb{R}^m$;
$|y|$ denotes the Euclidean norm of $y$;
$[v,z]$ denotes the line segment connecting the two points $v,z$;
line$(v,z)$ denotes the line passing the two points $v,z$;
for a set $K\subseteq\mathbb{R}^m$, int$(K)$ and
bd$(K)=K\backslash$int$(K)$ denote the sets of
interior points and boundary points of $K$, respectively;
for a closed convex set $K\subseteq\mathbb{R}^m$,
$P_K(\cdot)$ denotes the projection operator onto $K$;
$|y|_K:=|y-P_K(y)|$ denotes the distance between $y$ and $K$;
$\langle \cdot, \cdot\rangle$ denotes the Euclidean inner product in $\mathbb{R}^m$;
the angle between nonzero vectors $y$ and $z$ is denoted as $\angle(y,z)\in[0,\pi]$, where
$\cos \angle(y,z)=\langle y, z\rangle/(|y||z|)$; span$\{v_1,...,v_p\}$ (aff$\{v_1,...,v_p\}$) denotes the
subspace (affine hull) generated by vectors $v_1,...,v_p$.

\section{Preliminaries}

In this section, we give preliminaries on graph theory \cite{God},
convex analysis \cite{Roc}, the consensus model with disturbances \cite{shi3}.

\subsection{Graph Theory}

A multi-agent network can be described by a directed graph ${\mathcal{G}}=(\mathcal{V},\mathcal{E})$, where
$\mathcal{V}=\{1,2,...,n\}$ is the node (or agent) set and
$\mathcal{E}\subseteq \mathcal{V} \times \mathcal{V}$ the arc set with
the arc $(j,i)\in\mathcal{E}$ describing that node $i$ can receive the information
of node $j$.  Here $(i,i)\not\in\mathcal{E}$ for all $i$.
Let $\mathcal{N}_i=\{j\in\mathcal{V}|(j,i)\in\mathcal{E}\}$
be the set of neighbors of node $i$.
A path from node $i$ to node $j$ in $\mathcal{G}$
is a sequence $(i, i_1),(i_1, i_2),..., (i_p, j)$ of arcs in $\mathcal{E}$. Graph $\mathcal{G}$ is said to be strongly
connected if there exists a path from $i$ to $j$ for each
pair of nodes $i,j\in\mathcal{V}$.
Graph $\mathcal{G}$ is undirected
when $(j,i)\in\mathcal{E}$ if and only if $(i,j)\in\mathcal{E}$.

The communication over the network under consideration is switching and
characterized by a directed graph process $\mathcal{G}_{\sigma(t)}=(\mathcal{V},\mathcal{E}_{\sigma(t)}),t\geq
0$, with $\mathcal{E}_{\sigma(t)}$ the arc set of the
graph at time $t$. Here $\sigma: [0,\infty)\rightarrow
\mathcal{Q}$ is a piecewise constant function to describe the
time-varying graph process, where $\mathcal{Q}$ is the index set of all possible
graphs on $\mathcal{V}$. Let $\Delta:=\{t_k,k\geq0\}$ with $t_0=0$ denote the set of all switching moments of switching graph $\mathcal{G}_{\sigma}$. As usual, we assume there is a dwell time $\tau>0$ between two consecutive graph
switching moments, i.e., $t_{k+1}-t_k\geq \tau$ for all $k$.
The switching graph $\mathcal{G}_{\sigma}$ is uniformly jointly strongly connected (UJSC) if there exists $T>0$ such that
the union graph $(\mathcal{V},\cup_{t\leq s<t+T}\mathcal{E}(s))$ is strongly connected for $t\geq 0$.

\subsection{Convex Analysis}

A set $K\subseteq \mathbb{R}^m$ is convex if $\lambda z_1
+(1-\lambda)z_2\in K$ for any $z_1, z_2 \in K$ and $0<\lambda<1$.
 For a closed convex set $K$ in $\mathbb{R}^m$, we can associate with any
$z\in \mathbb{R}^m$ a unique element $P_K(z)\in K$ satisfying
$|z-P_K(z)|= \inf_{y\in K}|z-y|=:|z|_K$, where $P_K$
is called the projection operator onto $K$.
We have the following properties for the projection operator $P_K$.

\begin{lem}\label{pro}
Let $K$ be a closed convex set in $\mathbb{R}^m$. Then
\begin{align}
(i)\quad &\langle y-P_K(y), z-P_K(y)\rangle\leq 0\;\mbox{for any}\; y\; \mbox{and}\; z\in K;\nonumber\\
(ii)\quad &|P_K(y)-z|\leq|y-z|\;\mbox{ for any}\; y\in\mathbb{R}^m \;\mbox{and any}\; z\in K;\nonumber\\
(iii)\quad &\langle y-P_K(y),z-y\rangle\leq |y|_K(|z|_K-|y|_K)\;\;\mbox{for any}\;\; y\;\mbox{and}\;z;\nonumber\\
(iv)\quad& |P_K(y)-P_K(z)|\leq|y-z|\;\;\mbox{for any}\;\; y\;\mbox{and}\;z.\nonumber
\end{align}
\end{lem}

\emph{Proof.} (i) is an equivalent definition of convex projection; (ii) comes from
Lemma 1 (b) in \cite{nedic2}. We now show (iii).
First of all, $\langle y-P_K(y), P_K(z)-P_K(y)\rangle\leq 0$ by (i).
It is also clear that $\langle y-P_K(y), z-P_K(z)\rangle\leq |y|_K|z|_K$. Then
\begin{align}
\big\langle y-P_K(y),z-y\big\rangle&=\big\langle y-P_K(y),z-P_K(z)+P_K(z)-P_K(y)+P_K(y)-y\big\rangle\nonumber\\
&\leq |y|_K|z|_K-|y|^2_K.\nonumber
\end{align}
Thus, the inequality (iii) follows. (iv) is the standard non-expansive property.
\hfill$\square$

The following lemma characterizes the distance between convex sets
and their nonempty intersection, which can be found from Proposition
5.6.1 on page 72 
in \cite{bauu}.

\begin{lem}\label{dis}
Let $K_1,...,K_n$ be closed convex sets in $\mathbb{R}^m$. If $\bigcap^n_{i=1}int(K_i)\neq\emptyset$,
then for every bounded set $S$, there exists $\kappa_S>0$ such that
$$
|x|^2_{\bigcap^n_{i=1}K_i}\leq\kappa_S\max_{1\leq i\leq n}|x|^2_{K_i}, \forall x\in S.
$$
\end{lem}

The following lemma  can be found from Proposition
1 on page 24 in \cite{aub}.

\begin{lem}\label{aub}
Let $K$ be a closed convex set in $\mathbb{R}^m$. Then
$|x|^2_{K}$ is continuously differentiable and
$$\nabla |x|^2_{K}=2(x-P_{K}(x)).$$
\end{lem}

A function $\varphi(\cdot): \mathbb{R}^m\rightarrow \mathbb{R}$ is said to be convex if
$\varphi(\lambda z_1+(1-\lambda)z_2)\leq\lambda \varphi(z_1)
+(1-\lambda)\varphi(z_2)$ for any $z_1,z_2 \in \mathbb{R}^m$ and $0<\lambda<1$,
and it is $\ell$-strongly convex if $\varphi(\lambda z_1+(1-\lambda)z_2)\leq\lambda \varphi(z_1)
+(1-\lambda)\varphi(z_2)-\frac{1}{2}\ell\lambda(1-\lambda)|z_1-z_2|^2$ for any $z_1,z_2 \in \mathbb{R}^m$ and $0<\lambda<1$.
The following two inequalities hold
for a continuously differentiable convex and $\ell$-strongly convex function $\varphi$, respectively:
\begin{align}
\label{conve}\varphi(y)&\geq \varphi(x)+\langle y-x,\nabla\varphi(x) \rangle,\forall x,y\in \mathbb{R}^m,\\
\label{conve1}\varphi(y)&\geq \varphi(x)+\langle y-x,\nabla\varphi(x) \rangle+\frac{\ell}{2}|y-x|^2,\forall x,y\in \mathbb{R}^m.
\end{align}


The upper Dini derivative of function $g:
(a, b)\rightarrow \mathbb{R}$ at $t\in(a, b)$ is defined as
$$
D^+g(t)=\limsup_{s\rightarrow0^+}\frac{g(t+s)-g(t)}{s}.
$$
$g$ is non-increasing on $(a,b)$ if $D^+g(t)\leq0,\forall t\in(a,b)$. The
following result was shown in \cite{dan}.

\begin{lem}\label{max}
Let $g_i(t,x): \mathbb{R}\times \mathbb{R}^m\rightarrow \mathbb{R}$, $i=1,...,n$ be continuously differentiable
and $g(t,x)=\max_{1\leq i\leq n}g_i(t,x)$. Then $D^+g(t,x(t))=\max_{i\in
\mathcal{I}(t)}\dot g_i(t,x(t))$ with
$\mathcal{I}(t)=\big\{i|g_i(t,x(t))=g(t,x(t)),$ $1\leq i\leq n\big\}$.
\end{lem}

\subsection{Consensus}
Consider the following consensus model with disturbance $w_i$,
\begin{equation}
\label{con}\dot z_i(t)=\sum_{j\in\mathcal{N}_i(t)}(z_j(t)-z_i(t))+w_i(t),\;i=1,...,n,
\end{equation}
where the disturbance $w_i(t):[0,\infty)\rightarrow \mathbb{R}$ is continuous. System
(\ref{con}) has a continuous solution, which satisfies (\ref{con}) for almost all $t$
except at the switching moments of switching graph $\mathcal{G}_\sigma$.
Consensus is said to be achieved for system (\ref{con}) if for any
initial condition, $\lim_{t\rightarrow\infty}|z_i(t)-z_j(t)|=0$ for
all $1\leq i,j\leq n$.

The next two lemmas can be obtained from
the proofs of Theorem 4.2 and Proposition 4.10 in \cite{shi3},
respectively.

\begin{lem}
\label{lemcon}
If the switching graph $\mathcal{G}_{\sigma}$ is UJSC for system (\ref{con}), then there
exist $0<\beta<1$ and $B_0,B_1>0$ such that
\begin{align}
H((k+1)B_0)&\leq \beta H(kB_0)+B_1\int^{(k+1)B_0}_{kB_0}\max_{1\leq i\leq n}|w_i(t)|dt,\;\forall k\geq 0,\nonumber\\
H(t)&\leq H(kB_0)+B_1\int^{(k+1)B_0}_{kB_0}\max_{1\leq i\leq n}|w_i(t)|dt,\;
\forall t: kB_0\leq t<(k+1)B_0,\nonumber
\end{align}
where
$H(t)=\max_{1\leq i,j\leq n}|z_i(t)-z_j(t)|$.
\end{lem}

\begin{lem}
\label{lem5} Suppose the switching graph $\mathcal{G}_{\sigma}$ of system (\ref{con}) is UJSC and $\lim_{t\rightarrow\infty}w_i(t)=0$ for all $i$. Then
consensus is achieved for system (\ref{con}).
\end{lem}

In the following consensus analysis, we need to extend the standard Barbalat's Lemma to switching cases.
A function $g:[0,\infty)\rightarrow \mathbb{R}$ is uniformly continuous with respect to time intervals $(s_k,s_{k+1})$, $s\geq0$
with 
$s_{k+1}-s_k\geq \upsilon$ for some $\upsilon>0$, if, for any $\varepsilon>0$, there is $\delta>0$, which depends on $\{s_k\}_{k\geq0}$
and $g$, such that,
for any $k$ and $r_1,r_2$ with $s_k<r_1<r_2<s_{k+1}$, $|g(r_2)-g(r_1)|\leq\varepsilon$
when $|r_2-r_1|\leq\delta$.
We now introduce an extended Babalat's Lemma, 
which can be found in \cite{lou15}.

\begin{lem}\label{mod}
For a continuous function $g$, suppose $\lim_{t\rightarrow\infty}g(t)=g_0$ exists and $g$ is continuously differential on each interval $(s_k,s_{k+1})$, whose
derivative $\dot g$ is uniformly continuous with respect to time intervals $(s_k,s_{k+1})$ for $k\geq0$.
Then $\lim_{t\rightarrow\infty}\dot g(t)=0$.
\end{lem}

\section{Approximate Projection and Problem Formulation}

In this section, we introduce the distributed SDOP and the
distributed continuous-time approximate projected algorithm.

Consider a network of $n$ agents (or nodes) and
bounded closed convex sets $X_i\subseteq \mathbb{R}^m$ for $i=1,...,n$,
with $X_i$ only associated with (or known by) agent $i$. The goal of the network is to cooperatively find a point $x^*$
with the shortest quadratic distance from the $n$ closed convex sets:
\begin{align}\label{op}
x^*\in\arg\min f(x),\quad f(x)=\sum^n_{i=1}|x|^2_{X_i}.
\end{align}


Projection-based methods have been widely adopted in the literature
to solve CIP and constrained optimization problems, and almost all methods
require that the exact projection can be obtained
\cite{bauu,shi1,shi2,nedic2,lin,joh09,meng}.  Since the exact projection may
be difficult to obtain in practice,
each agent may only obtain an approximate projection point located on the convex set surface and near the exact projection point.
To be strict, we give the following definition.

\begin{defn}
\label{def}
Let $0\leq\theta<\pi/2$ and $K$ be a closed convex
set in $\mathbb{R}^m$.
Define sets
\begin{align}
&\mathbf{C}_K(v,\theta)=v+\big\{ z|\;\langle z, P_K(v)-v\rangle\geq |z||v|_K\cos\theta \big\},\nonumber\\
&\mathbf{b}(v,K)=\Big\{z\big|z\in bd(K),[v,z]\cap bd(K)=\{z\}\Big\}.\nonumber
\end{align}
The approximate projection $\mathbf{P}^a_K(v,\theta)$
of point $v$ onto $K$ is defined as the following set:
$$
\mathbf{P}^a_K(v,\theta)=
\left\{
\begin{array}{ll}
\mathbf{C}_K(v,\theta)\cap \mathbf{b}(v,K),\qquad\;\;\;\;\mbox{if}\;\;v\not\in K;\\
\{v\},\qquad\qquad\qquad\qquad\;\;\;\;\;\; \mbox{otherwise.}\\
\end{array}
\right.
$$
\end{defn}

\begin{figure}[!htbp]
\centering
\includegraphics[width=2.2in]{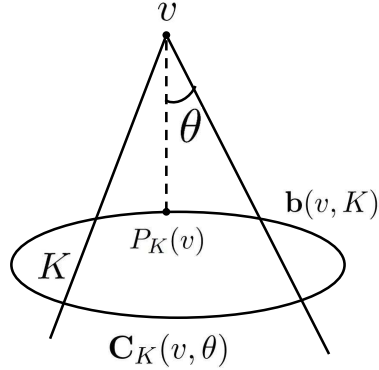}
\caption{The approximate projection of point $v$ onto closed convex set $K$.}\label{fi}
\end{figure}

As shown in Fig. \ref{fi}, the cone $\mathbf{C}_K(v,\theta)-v$ consists of all vectors having
angle with the direction $P_K(v)-v$ less than $\theta$, and
$\mathbf{b}(v,K)$ is the region on the boundary of $K$ that the agent can ``see" starting from point $v$.
Obviously, the exact projection
$P_K(v)\in\mathbf{P}^a_{K}(v,\theta)$ for any $v\in \mathbb{R}^m$
and $0\leq\theta<\pi/2$ and $\mathbf{P}^a_{K}(v,0)=\{P_K(v)\}$.

\begin{rem}
The approximate projection is more ``practical" than the exact projection.   For example,
a robot likes to get its exact projection point on its convex target set when it approaches the set.  However, in reality, it may select another point on the set surface as the exact one by mistake or to avoid expensive measurement or tedious computation.   Then the selected projection point becomes an approximate one.  In other words, this concept captures the
situation when agents can only obtain some point on the set surface, which may not be but close to the exact projection point.  Note that this concept is different from that given in \cite{lou12}, where the approximate projection point is located in a ``triangle'' region
specified by $v$, the hyperplane of $K$ on $P_K(v)$
and the approximate angle $\theta$.
\end{rem}


We next give some basic assumptions for our following analysis.

\noindent {\bf A1} (Connectivity)  The switching graph $\mathcal{G}_{\sigma}$ is UJSC.

\noindent {\bf A2} (Convex Sets)  (i) The boundary surfaces of convex sets $X_i,i=1,...,n$ are regular (or smooth);

(ii) The convex set $X_i$ contains nonempty interior points for $i=1,...,n$.

The definition of regularity or smoothness of a manifold can be easily found (referring to Definition 1 on page 52 in \cite{com} for more details).
Note that the Gaussian curvature of regular (or smooth) surfaces of closed bounded sets are bounded.
In fact, {\bf A2} is quite mild.  The boundaries of many well-known sets, such as the surfaces of spheres, ellipsoids, are regular; and moreover, the assumption that set $X_i\subseteq\mathbb{R}^m$ contains nonempty interior points is equivalent to dim$(X_i)=m$ (where dim$(X_i)$ denotes the dimension of the affine hull of set $X_i$), which was also widely used in the literature.

Let $P^a_{X_i}(\cdot):\mathbb{R}^m\rightarrow\mathbb{R}^m$ be a continuous map with $P^a_{X_i}(v)\in \mathbf{P}^a_{X_i}(v,\theta_i(v))$ for any $v$, where
$$\theta_i(v)=\angle(P^a_{X_i}(v)-v,P_{X_i}(v)-v),$$
$0\leq\theta_i(v)<\pi/2$. Let $\theta_i(v)=0$ for simplicity when $v\in X_i$.
In this paper, $\theta_i(v)$ is referred to as the approximate angle of $v$ onto $X_i$.
The following assumption was used in Lou et al. (2014).

\noindent {\bf A3} (Approximate Angle)
There exists $0<\theta^*<\pi/2$ such that
$0\leq\theta_i(v)\leq \theta^*$ for all $i,v$.

Here we propose a distributed continuous-time approximate projected algorithm:
\begin{equation}
\label{algo0}\dot x_i(t)=\sum_{j\in\mathcal{N}_i(t)}(x_j(t)-x_i(t))+\alpha_t(P^a_{X_i}(x_i(t))-x_i(t)),\;i=1,...,n,
\end{equation}
where $x_i\in \mathbb{R}^m$ is the state estimate of agent $i$ for the optimal solution, $\mathcal{N}_i(t)$ is the neighbor set of node $i$ at time $t$, $\{\alpha_t\}$ is the stepsize $(0\leq \alpha_{t}\leq \alpha^*,\alpha^*>0)$ and is uniformly continuous over $t$.
The continuity of stepsize $\alpha_t$ and maps $P^a_{X_i}(\cdot)$ guarantees that
(\ref{algo0}) has a solution that is continuous over $[0,\infty)$
and continuously differentiable except at the switching moments
of switching graph $\mathcal{G}_\sigma$.

\begin{rem}
The term $P^a_{X_i}(x)-x$ can be viewed as a negative ``approximate" gradient because it becomes the negative gradient of $\frac{1}{2}|x|^2_{X_i}$ by noting that $\nabla|x|^2_{X_i}=2(x-P_{X_i}(x))$ in the exact projection case (i.e., $P^a_{X_i}(x_i(t))=P_{X_i}(x_i(t))$).
In fact, (\ref{algo0}) with taking $\alpha_{t}\equiv 1$ and exact projection
was proposed in \cite{shi2} to solve the CIP.
\end{rem}

\begin{figure}[!htbp]
\centering
\includegraphics[width=2.6in]{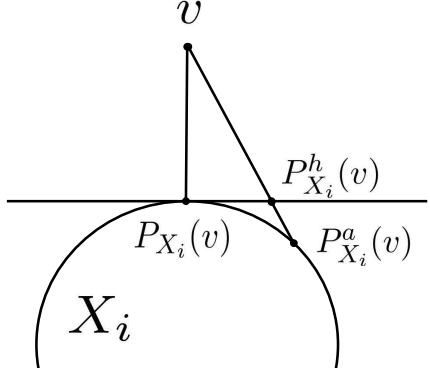}
\caption{An illustration for $P^h_{X_i}(v)$.}\label{f22}
\end{figure}

The convergence analysis of (\ref{algo0}) is not easy because the gradient term is corrupted with state-dependent approximation and there is no explicit expression to describe the relationship between the approximate projection point and the exact one. To handle the problem, we make some transformation.
In the case of $v\not\in X_i$, we define by $P^h_{X_i}(v)$ the intersection point of the hyperplane of $X_i$ at $P_{X_i}(v)$ (the tangent plane of bd$(X_i)$ at $P_{X_i}(v)$) with $P_{X_i}(v)-v$ as the normal direction and the line segment $[v,P^a_{X_i}(v)]$, as shown in Fig. \ref{f22}.  Clearly, $P^h_{X_i}(v)=P_{X_i}(v)$ when $P^a_{X_i}(v)=P_{X_i}(v)$. In the case of $v\in X_i$, we define
$P^h_{X_i}(v)=P^a_{X_i}(v)=v$. Then we can find that $P^h_{X_i}(v)=v$ if and only if $v\in X_i$.
We write
$$
P^a_{X_i}(v)-v=\gamma_{X_i}(v)(P^h_{X_i}(v)-v),
$$
where $\gamma_{X_i}(v)=\frac{|P^a_{X_i}(v)-v|}{|P^h_{X_i}(v)-v|}\geq 1$ if $P^h_{X_i}(v)\neq v$,
and $\gamma_{X_i}(v)=1$ otherwise.

Rewrite $\alpha_t(P^a_{X_i}(x_i(t))-x_i(t))=\alpha_{i,t}(P^h_{X_i}(x_i(t))-x_i(t))$,
with the virtual stepsize of agent $i$ defined as
$$
\alpha_{i,t}=\left\{
  \begin{array}{ll}
    \gamma_{X_i}(x_i(t))\alpha_t=\frac{|P^a_{X_i}(x_i(t))-x_i(t)|}{|P^h_{X_i}(x_i(t))-x_i(t)|}\alpha_t, & \hbox{\mbox{if}\;$P^h_{X_i}(x_i(t))\neq x_i(t)$;} \\
    \alpha_t, & \hbox{otherwise.}
  \end{array}
\right.
$$
Clearly, $\alpha_{i,t}\geq\alpha_t$.  Although the designed stepsize $\alpha_t$ is the same for all agents, agent $i$ has its own virtual stepsize $\alpha_{i,t}$ based on its own approximate projection.
We express (\ref{algo0}) in another form with heterogeneous virtual stepsizes:
\begin{align}\label{system}
\dot x_i(t)=\sum_{j\in\mathcal{N}_i(t)}(x_j(t)-x_i(t))+\alpha_{i,t}(P^h_{X_i}(x_i(t))-x_i(t)), i=1,...,n.
\end{align}

Then we give a definition for our problem as follows.

\begin{defn}
\label{def1} The shortest distance optimization problem (SDOP) is solved by (\ref{algo0}) or (\ref{system})
if, for any initial condition $x_i(0)\in \mathbb{R}^m,i=1,...,n$,
there exists $x^*\in \arg\min\sum^n_{i=1}|x|^2_{X_i}$ such that
$$
\lim_{t\rightarrow\infty}x_i(t)=x^*,\;i=1,...,n.\;
$$
\end{defn}

In the following three sections, we first establish some basic results, and then present the convergence results in the nonempty intersection and empty intersection cases.

\section{Discussions on Boundedness and Stepsizes}

In this section, we show the state boundedness and
establish an ``equivalent" relationship between the
designed stepsize $\alpha_t$ and the virtual stepsize $\alpha_{i,t}$.

\subsection{Boundedness of System States}

Denote $\theta_{i,t}=\theta_i(x_i(t))$ for simplicity.
Note that $\theta_{i,t}$ is also equal to $\angle(P^h_{X_i}(x_i(t))-x_i(t),P_{X_i}(x_i(t))-x_i(t))$.
Here we study the boundedness of $x_i(t),i\in \mathcal{V},t\geq0$ of (\ref{system}) with the approximate angle $\theta_{i,t}$.

Let $X_{c}=\mbox{co}\{X_i,i=1,...,n\}$ be the convex hull
of the sets $X_i,i=1,...,n$, $\xi:=\sup_{z_1,z_2\in X_{c}}|z_1-z_2|$, which is finite due to the boundedness of $X_i$s.

 \begin{them}
\label{bounded}
(i) For any initial condition $x_i(0),i\in\mathcal{V}$, the system states $x_i(t),i\in \mathcal{V},t\geq0$ are bounded;

(ii) Suppose $\liminf_{t\rightarrow\infty}\alpha_t>0$. Then, for any $0<\theta<\pi/2$
and any initial condition $x_i(0),i\in\mathcal{V}$,
$$
\limsup_{t\rightarrow\infty}|x_i(t)|_{X_{c}}\leq\max\Big\{\frac{\xi}{\sin\theta},\xi\Big(\tan\theta+\sqrt{(\tan\theta)^2+2\tan\theta}\Big)\Big\}.
$$
Furthermore, if {\bf A3} holds,
then for any initial condition $x_i(0),i\in\mathcal{V}$,
$$\limsup_{t\rightarrow\infty}|x_i(t)|_{X_{c}}\leq\xi\big(\tan\theta^*+\sqrt{(\tan\theta^*)^2+2\tan\theta^*}\big).$$
\end{them}

\begin{rem}\label{rem1}
A discrete-time algorithm was
proposed in \cite{lou12} to solve CIP with approximate projection, where $\pi/4$ was found to be
a critical approximate angle ensuring the boundedness of system states in the
case of $\alpha_{i,k}\equiv1$ and $\theta_{i,k}\equiv\theta$, $0\leq\theta<\pi/2$.
To be specific, the states are uniformly bounded with respect to all initial conditions when $\theta<\pi/4$
and unbounded for most all initial conditions when $\theta>\pi/4$.
Different from this critical approximate angle result, Theorem \ref{bounded} shows that the continuous-time system states
are always bounded for any initial
condition no matter how large $\theta$ is, and moreover, the states are
uniformly bounded for all initial conditions with fixing $\alpha_{i,t}\equiv1$.
\end{rem}

\begin{rem}
Notice that the boundedness results in Theorem \ref{bounded}
do not require any connectivity of communication graph.
Moreover, when the exact projection is obtained ($\theta_{i,t}=0$),
Theorem \ref{bounded} (ii) implies that all agents converge to the convex hull spanned by all the convex sets, which is related to
the target aggregation and leader-following problems \cite{louhong,hu,shi09}.
\end{rem}

Here we present a result for a simple case: there is only one node in the
network. Its set is bounded and denoted as $X_\ast$. We denote the
states of this node as $x_\ast(t),t\geq0$ driven by the
continuous-time approximate projected dynamical system:
$$
\dot x_*(t)=\alpha_t(P^a_{X_*}(x_*(t))-x_*(t)),
$$
where $P^a_{X_*}(x_*(t))\in\mathbf{P}^a_{X_*}(x_*(t),\theta_*(x_*(t)))$.
Noticing that $\big\langle x_*(t)-P_{X_*}(x_*(t)), P^a_{X_*}(x_*(t))-P_{X_*}(x_*(t))\big\rangle\leq0$,
we have
$$
\frac{d|x_*(t)|^2_{X_*}}{dt}=2\big\langle x_*(t)-P_{X_*}(x_*(t)), \dot x_*(t)\big\rangle
\leq-2\alpha_t |x_*(t)|^2_{X_*}.
$$
Then for any initial condition $x_*(0)$ and any stepsize $\{\alpha_t\}$,
$|x_*(t)|_{X_*}\leq|x_*(0)|_{X_*}$ always holds.

Now we present the proof of Theorem \ref{bounded}.

\emph{Proof.}
Let $t\not\in\Delta$.
Denote $\hbar_i(t)=\frac{1}{2}|x_i(t)|^2_{X_{c}}$, $\hbar(t)=\max_{1\leq i\leq n}\hbar_i(t)$.
By Lemmas \ref{aub} and \ref{max}, we have
\begin{align}
\label{a42}D^+\hbar(t)&=\max_{i\in \mathcal{I}(t)}\langle x_i(t)-P_{X_{c}}(x_i(t)),\dot x_i(t)\rangle\nonumber\\
&=\max_{i\in \mathcal{I}(t)}\big\langle x_i(t)-P_{X_{c}}(x_i(t)),\sum_{j\in\mathcal{N}_{i}(t)}(x_j(t)- x_i(t))+\alpha_{i,t}(P^h_{X_i}(x_i(t))-x_i(t))\big\rangle,
\end{align}
where $\mathcal{I}(t)=\big\{i|i\in\mathcal{V},\hbar_i(t)=\hbar(t)\big\}$. Take $i\in\mathcal{I}(t)$.
Lemma \ref{pro} (iii) implies that, for any $j$,
\begin{align}\label{a43}
\langle x_i-P_{X_{c}}(x_i),x_j-x_i\rangle&\leq |x_i|_{X_c}(|x_j|_{X_{c}}-|x_i|_{X_{c}})\leq0.
\end{align}
According to Lemma \ref{pro} (i),
$\langle x_i-P_{X_{c}}(x_i),P_{X_i}(x_i)-P_{X_{c}}(x_i)\rangle\leq0$ due to $X_i\subseteq X_{c}$. Therefore,
\begin{align}\label{a40}
\langle x_i-P_{X_{c}}(x_i),P_{X_i}(x_i)-x_i\rangle\leq -|x_i|^2_{X_{c}}.
\end{align}
Moreover, recalling the definitions of $P^h_{X_i}(x_i(t))$ and $\theta_{i,t}$, we have $\langle x_i(t)-P_{X_i}(x_i(t)),P^h_{X_i}(x_i(t))-P_{X_i}(x_i(t))\rangle=0$
and $|P^h_{X_i}(x_i(t))-P_{X_i}(x_i(t))|=\tan\theta_{i,t}|x_i(t)|_{X_i}$. Then
\begin{align}\label{a41}
&\ \ \langle x_i(t)-P_{X_{c}}(x_i(t)),P^h_{X_i}(x_i(t))-P_{X_i}(x_i(t))\rangle\nonumber\\
&=\langle x_i(t)-P_{X_i}(x_i(t))+P_{X_i}(x_i(t))-P_{X_c}(x_i(t)),P^h_{X_i}(x_i(t))-P_{X_i}(x_i(t))\rangle\nonumber\\
&\leq|P_{X_i}(x_i(t))-P_{X_{c}}(x_i(t))||P^h_{X_i}(x_i(t))-P_{X_i}(x_i(t))|\nonumber\\
&\leq\xi\tan\theta_{i,t}|x_i(t)|_{X_i}\nonumber\\
&\leq\xi\tan\theta_{i,t}(|x_i(t)|_{X_{c}}+\xi),
\end{align}
where the last inequality follows from
the relation $|x_i|_{X_i}\leq|x_i-P_{X_{c}}(x_i)|+|P_{X_{c}}(x_i)-P_{X_i}(x_i)|\leq|x_i|_{X_{c}}+\xi$.
Thus, based on (\ref{a40}) and (\ref{a41}), we have
\begin{align}\label{a44}
&\langle x_i(t)-P_{X_{c}}(x_i(t)),P^h_{X_i}(x_i(t))-x_i(t)\rangle\leq -|x_i(t)|^2_{X_{c}}+\xi\tan\theta_{i,t}(|x_i(t)|_{X_{c}}+\xi).
\end{align}

With (\ref{a42}), (\ref{a43}), (\ref{a44}) and $i\in\mathcal{I}(t)$, we obtain
$$
D^+\hbar(t)\leq\alpha_{i,t}(-2\hbar(t)+\xi\tan\theta_{i,t}(\sqrt{2\hbar(t)}+\xi)).
$$
We complete the proof by the following analysis.

(i) It is easy to see that for any $0<\hat\theta<\pi/2$,
$\theta_{i,t}\leq\hat\theta$ when $|x_i(t)|_{X_c}\geq\xi/\sin\hat\theta$.
Then $D^+\hbar(t)\leq0$ when $t\not\in\Delta$ and
$$\hbar(t)\geq\max\Bigg\{\frac{\xi^2}{2(\sin\hat\theta)^2},\frac{\xi^2(\tan\hat\theta)^2}{4}
+\frac{\xi^2\tan\hat\theta}{2}\Big(1+\frac{\sqrt{(\tan\hat\theta)^2+4\tan\hat\theta}}{2}\Big)\Bigg\}.$$
Hence, system states are bounded for any initial condition.

(ii)
Let $\alpha_*=(\liminf_{t\rightarrow\infty}\alpha_t)/2>0$
and $\hat t$ be the moment such that when $t\geq\hat t$,
$\alpha_{i,t}\geq\alpha_t\geq \alpha_*$.
Then $D^+\hbar(t)\leq-\alpha_*\hbar(t)$ once $t\geq \hat t$, $t\not\in\Delta$ and
$$\hbar(t)\geq\max\Bigg\{\frac{\xi^2}{2(\sin\theta)^2},\xi^2(\tan\theta)^2
+\xi^2\tan\theta\Big(1+\sqrt{(\tan\theta)^2+2\tan\theta}\Big)\Bigg\}.$$
Therefore, $\hbar(t)$ is not greater than the number in the preceding inequality when $t\geq \hat t$ and $t\not\in\Delta$.
The second conclusion can be obtained directly based on the above arguments.

Thus,
the conclusion follows from the continuity of $x_i(t)$. \hfill$\square$

\subsection{Equivalence between Stepsizes}

To obtain the convergence conditions, we establish a relationship between the designed stepsize $\alpha_t$
and virtual stepsizes $\alpha_{i,t}$.  To show that they are equivalent in the sense that they can be bounded by each other, it suffices to establish the boundedness of $\gamma_{X_i}(\cdot)$.

Let $S=X_c+\mathbf{B}(0,r_0)$,
where $\mathbf{B}(0,r_0)$ denotes the ball with center zero and radius $r_0>0$.
Denote
$\mu_i(v)=\angle(P_{X_i}(v)-P^a_{X_i}(v),v-P^a_{X_i}(v))$.

\begin{lem}\label{props}
If the map $P^a_{X_i}(\cdot)$ satisfies
\begin{align}\label{condition}
\inf_{v\in S\backslash X_i,P^a_{X_i}(v)\neq P_{X_i}(v)}\mu_i(v)>0,
\end{align}
then $\sup_{v\in S}\gamma_{X_i}(v)<\infty$.

\end{lem}

Its proof is in the Appendix.
The following result provides a condition to guarantee the condition (\ref{condition}).

\begin{lem}\label{props2}
Suppose {\bf A2} and {\bf A3} hold. Then (\ref{condition}) holds.
\end{lem}

The proof is also in the Appendix.
Because the states of (\ref{system}) are bounded by Theorem \ref{bounded}, we take
sufficiently large $r_0$ such that
$S$ contains all the system states.
We next show that {\bf A2} and {\bf A3} imply the equivalence between
$\{\alpha_t\}$ and $\{\alpha_{i,t}\}$.

Clearly, $\alpha_t=\alpha_{i,t}$ when $x_i(t)\in X_i$ or $P^a_{X_i}(x_i(t))=P_{X_i}(x_i(t))$.
Then we only need to focus on the case when $x_i(t)\not\in X_i$ and
$P^a_{X_i}(x_i(t))\not=P_{X_i}(x_i(t))$.
By (\ref{ef}) in the Appendix,
$\gamma_{X_i}(x_i(t))\leq 1+\frac{1}{\sin\mu_{i,t}}\sin \theta_{i,t}$,
where $\mu_{i,t}:=\mu_i(x_i(t))$.
Then we have

\begin{them}
\label{thnew}
Under {\bf A2} and {\bf A3},
\begin{align}\label{rg}
\alpha_t\leq \alpha_{i,t}\leq C_{i,t}\alpha_t\leq C_{i}\alpha_t,\;\forall t,
\end{align}
where $C_{i,t}=1+\frac{1}{\sin\mu_{i}}\sin \theta_{i,t}$, $C_i=1+\frac{1}{\sin\mu_{i}}$,
\begin{align}\label{rgg}
\mu_{i}&:=\inf_{t\geq0,x_i(t)\not\in X_i,P^a_{X_i}(x_i(t))\not=P_{X_i}(x_i(t))}\mu_{i,t}\nonumber\\
&\geq\inf_{v\in S\backslash X_i,P^a_{X_i}(v)\neq P_{X_i}(v)}\mu_i(v)>0.
\end{align}
\end{them}

Note that the first inequality of (\ref{rgg}) follows from $x_i(t)\in S$
and the second one from Lemma \ref{props2}.
In fact, (\ref{rg}) somehow characterizes the bounded bending property
of smooth surfaces, which helps convert the convergence conditions on $\alpha_{i,t}$ into the conditions on $\alpha_t$.

\begin{rem}
As Theorem \ref{thnew} shows, {\bf A2} and {\bf A3} guarantee the equivalence between the designed stepsize and the virtual stepsize.
In fact, with (\ref{rg}), we found that
under {\bf A1}, the optimal convergence established in the next two sections hold for general convex sets (not necessary to satisfy {\bf A2} and {\bf A3}).
\end{rem}

\section{Nonempty Intersection Case}

In this section, we show the convergence result in the nonempty intersection case,
$\cap^n_{i=1}X_i\neq \emptyset$.
Clearly, $X_0:=\cap^n_{i=1}X_i$
is the optimal solution set of $\min\sum^n_{i=1}|x|^2_{X_i}$.

\begin{them}
\label{thm2} Suppose {\bf A1}-{\bf A3} hold and
$\bigcap^n_{i=1}X_i\neq\emptyset.$ Then SDOP is solved by system (\ref{system}) if
$\int^{\infty}_{0}\alpha_tdt=\infty$,
$\int^{\infty}_{0}\alpha_t\tan\theta^+_tdt<\infty$. Furthermore, in the special case when $\theta_{i,t}=0\;\forall i,t$, SDOP is solved by (\ref{system}) if and only if
$\int^{\infty}_{0}\alpha_tdt=\infty$.
\end{them}

\begin{rem}
When the intersection set of all $X_i$s is
nonempty, SDOP (\ref{op}) is
equivalent to CIP of finding
a point in $X_0$ \cite{inter3,inter2,lou13,lou12,shi1,shi2,nedic2,meng}. The
optimal consensus algorithm based on the exact projection presented
in \cite{shi2} is a special case of (\ref{algo0}) with taking
$\alpha_{t}\equiv1$ and $\theta_{i,t}\equiv0$, which
is consistent with Theorem \ref{thm2}. Theorem \ref{thm2} is also consistent with the convex intersection computation results of discrete-time algorithms in \cite{lou12,nedic2,meng}.
\end{rem}

Denote
$\alpha^+_t=\max_{1\leq i\leq n}\alpha_{i,t}$, $\theta^+_t=\max_{1\leq i\leq n}\theta_{i,t}$,
and the distance functions
$$h(t)=\max_{1\leq i\leq n}h_i(t),\;\;h_i(t)=\frac{1}{2}|x_i(t)|^2_{X_0},\;i=1,...,n,\;t\geq 0.$$

\begin{lem}\label{lemb2}
Suppose $\cap^n_{i=1}X_i\neq \emptyset$.
Then $D^+h(t)\leq2\alpha^+_{t}\tan\theta^+_{t}h(t)$ for any $t\not\in \Delta$.
\end{lem}

\emph{Proof.}  Let $t\not\in\Delta$.
Similar to (\ref{a42}), we have
\begin{align}
\label{a14}D^+h(t)=\max_{i\in \mathcal{I}(t)}\big\langle x_i(t)-P_{X_0}(x_i(t)),\sum_{j\in\mathcal{N}_{i}(t)}(x_j(t)- x_i(t))+\alpha_{i,t}(P^h_{X_i}(x_i(t))-x_i(t))\big\rangle,
\end{align}
where $\mathcal{I}(t)=\big\{i|i\in\mathcal{V},h_i(t)=h(t)\big\}$. Take $i\in\mathcal{I}(t)$.
Similar to (\ref{a43}), we also have
\begin{align}
\label{pq}\big\langle x_i(t)-P_{X_0}(x_i(t)),x_j(t)-x_i(t)\big\rangle\leq |x_i(t)|_{X_0}\big(|x_j(t)|_{X_0}-|x_i(t)|_{X_0}\big)\leq0.
\end{align}

From
$\big\langle x_i(t)-P_{X_i}(x_i(t)),P^h_{X_i}(x_i(t))-P_{X_i}(x_i(t))\big\rangle=0$, we have
\begin{align}\label{a16}
&\big\langle x_i(t)-P_{X_0}(x_i(t)),P^h_{X_i}(x_i(t))-P_{X_i}(x_i(t))\big\rangle\nonumber\\
&=\big\langle P_{X_i}(x_i(t))-P_{X_0}(x_i(t)),P^h_{X_i}(x_i(t))-P_{X_i}(x_i(t))\big\rangle\nonumber\\
&\leq |x_i(t)|_{X_0}\tan \theta_{i,t}|x_i(t)|_{X_i}\nonumber\\
&\leq \tan \theta_{i,t}|x_i(t)|^2_{X_0},
\end{align}
where the inequalities follow from Lemma \ref{pro} (ii) by setting
$K=X_i,y=x_i(t),z=P_{X_0}(x_i(t))\in X_i$, and $|x_i(t)|_{X_i}\leq|x_i(t)|_{X_0}$ (due to $X_0\subseteq X_i$).
Moreover, it follows from Lemma \ref{pro} (i) that
$\big\langle P_{X_i}(x_i(t))-P_{X_0}(x_i(t)),P_{X_i}(x_i(t))-x_i(t)\big\rangle\leq0$ and then
\begin{align}\label{a23}
&\big\langle x_i(t)-P_{X_0}(x_i(t)),P_{X_i}(x_i(t))-x_i(t)\big\rangle\nonumber\\
&=-|x_i(t)|^2_{X_i}+\big\langle P_{X_i}(x_i(t))-P_{X_0}(x_i(t)),P_{X_i}(x_i(t))-x_i(t)\big\rangle\nonumber\\
&\leq-|x_i(t)|^2_{X_i}.
\end{align}
Therefore, based on (\ref{a16}) and (\ref{a23}) we have
\begin{align}
&\big\langle x_i(t)-P_{X_0}(x_i(t)),P^h_{X_i}(x_i(t))-x_i(t)\big\rangle\nonumber\\
&=\big\langle x_i(t)-P_{X_0}(x_i(t)),P_{X_i}(x_i(t))-x_i(t)\big\rangle
+\big\langle x_i(t)-P_{X_0}(x_i(t)),P^h_{X_i}(x_i(t))-P_{X_i}(x_i(t))\big\rangle\nonumber\\
&\leq -|x_i(t)|^2_{X_i}+\tan \theta_{i,t}|x_i(t)|^2_{X_0}\nonumber\\
\label{a17}
&\leq\tan \theta_{i,t}|x_i(t)|^2_{X_0}.
\end{align}

From (\ref{a14}), (\ref{pq}) and (\ref{a17}),
$D^+h(t)\leq2\alpha_{i,t}\tan\theta_{i,t}h(t)\leq2\alpha^+_{t}\tan\theta^+_th(t).$
Thus, the conclusion follows. \hfill$\square$

\begin{lem}
\label{bound} If $\cap^n_{i=1}X_i\neq \emptyset$ and $\int^{\infty}_{0}\alpha^+_t\tan\theta^+_tdt<\infty$,
then $\lim_{t\rightarrow\infty}h(t)$ is a finite number.
\end{lem}

\emph{Proof.}
Lemma \ref{lemb2} implies
$h(t)\leq e^{2\int^\infty_{0}\alpha^+_s\tan\theta^+_sds}h(0)$
and then $\bar h:=\sup_{t\geq0}h(t)$ is a finite number. We then show the conclusion by contradiction.
Hence suppose there are two limit points $\bar h_1\neq \bar h_2$ of $\{h(t)\}_{t\geq0}$ such that
$\lim_{k\rightarrow\infty}h(s^1_k)=\bar h_1$ and
$\lim_{k\rightarrow\infty}h(s^2_k)=\bar h_2$. Without loss of
generality, we assume $\bar h_1<\bar h_2$. Clearly, for any
$\varepsilon>0$ for which $(1+\varepsilon)(\bar h_1+\varepsilon)
\leq \frac{\bar h_1+\bar h_2}{2}$, there is an integer $T_0>0$ such
that $e^{2\int^\infty_{s^1_k}\alpha^+_s\tan\theta^+_sds}\leq
1+\varepsilon$ and $|h(s^1_k)-\bar h_1|\leq \varepsilon$ for $k\geq
T_0$. According to Lemma
\ref{lemb2}, for any $t\geq s^1_k$ with $k\geq T_0$,
$$
h(t)\leq e^{2\int^\infty_{s^1_k}\alpha^+_s\tan\theta^+_sds}h(s^1_k)
\leq (1+\varepsilon)(\bar h_1+\varepsilon)
\leq \frac{\bar h_1+\bar h_2}{2}<\bar h_2,
$$
which contradicts that $\bar h_2$ is also a limit point of $\{h(t)\}_{t\geq0}$.
Thus, the conclusion follows.
\hfill$\square$

By Lemma \ref{bound}, if $\int^{\infty}_{0}\alpha^+_{t}\tan\theta^+_tdt<\infty$,
then the sequence $\{h(t)\}_{t\geq0}$ converges to a finite number denoted as $h^*$,
$$
\lim_{t\rightarrow\infty}h(t)=h^*.
$$
Denote
$h^+_i=\limsup_{t\rightarrow\infty}h_i(t),h^-_i=\liminf_{t\to
\infty}h_i(t),i\in\mathcal{V}.$
Clearly, $0\leq h^-_i\leq h^+_i\leq h^*$ for all $i$.
\begin{lem}
\label{lemA2} Suppose  {\bf A1}  holds and $\cap^n_{i=1}X_i\neq \emptyset$.
If $\int^{\infty}_{0}\alpha^+_{t}\tan\theta^+_tdt<\infty$ and there
exists some node $i_0\in\mathcal{V}$ such
that $h^-_{i_0}<h^*$, then $h^*=0$.
\end{lem}

The proof of Lemma \ref{lemA2} can be completed with similar arguments in Lemma 4.3 in \cite{shi2}, which is omitted here.

Now it is time to prove Theorem \ref{thm2}.

\emph{Proof of Theorem \ref{thm2}.}
Denote $\alpha^+_t=\max_{1\leq i\leq n}\alpha_{i,t}$.
From (\ref{rg}),  we find that $\int^{\infty}_{0}\alpha_tdt=\infty$,
$\int^{\infty}_{0}\alpha_t\tan\theta^+_tdt<\infty$ are equivalent to $\int^{\infty}_{0}\alpha^+_tdt=\infty$,
$\int^{\infty}_{0}\alpha^+_t\tan\theta^+_tdt<\infty$, respectively.

 Based on the similar arguments in
Lemmas \ref{lemb2} and \ref{bound}, we can show that, for any $z\in X_0$, the limit
$\lim_{t\rightarrow\infty}\max_{1\leq i\leq n}|x_i(t)-z|^2$ exists. Therefore, if consensus is achieved and $h^*=0$,
all agents will converge to a common point in $X_0$. Thus, it suffices to show that
consensus is achieved and $h^*=0$.

Because
\begin{align}\label{relation}
|P^h_{X_i}(x_i(t))-x_i(t)|=\frac{|x_i(t)|_{X_i}}{\cos\theta_{i,t}}\leq \frac{\sqrt{2h(t)}}{\cos\theta^*},
\end{align}
it follows that, if $h^*=0$, the second term on the right-hand side of
(\ref{system}) tends to zero as $t\to\infty$ and then the consensus is achieved for system (\ref{system}) by Lemma \ref{lem5}.
Therefore, it suffices to show $h^*=0$ in what follows.

In fact, if there is some node $i_0$ with $h^-_{i_0}<h^*$, then $h^*=0$
from the previous statements. Therefore, we need to prove $h^*=0$ from $h^+_{i}=h^-_{i}=h^*,\forall i$ by contradiction.
Clearly, for any $\varepsilon>0$, there is $\bar t>0$ such that when $t\geq \bar t$,
$|x_i(t)|_{X_0}\leq \sqrt{2h^*}+\varepsilon=:\phi$. We complete the proof by the following two steps.

Step (i). Suppose $h^+_{i}=h^-_{i}=h^*>0,\forall i$.
We claim that consensus can be achieved for system (\ref{system}).

We first show that $\lim_{t\rightarrow\infty}\alpha_{i,t}|x_i(t)|^2_{X_i}=0$ by contradiction.
Hence suppose there exist $i_0$ and an increasing subsequence $\{s_k\}_{k\geq0}$
with $\lim_{k\rightarrow\infty}s_k=\infty$ such that
$\alpha_{i_0,s_k}|x_{i_0}(s_k)|^2_{X_{i_0}}\geq c$ for some $c>0$.
Without loss of generality, we assume $s_0$ is sufficiently large such that
 $s_0\geq \bar t$ and $\int^\infty_{s_0}\alpha^+_t\tan \theta^+_tdt\leq \varepsilon/\sqrt{2h^*}$.

From Lemma \ref{aub}, the boundedness of system states and
(\ref{relation}), we know that $|x_{i_0}(t)|^2_{X_{i_0}}$ is uniformly continuous on $[0,\infty)$.
This along with the uniform continuity of $\alpha_t$ again implies that  $\alpha_t|x_{i_0}(t)|^2_{X_{i_0}}$ is also uniformly continuous on $[0,\infty)$.
Therefore, there is $\delta>0$ such that $\alpha_{i_0,t}|x_{i_0}(t)|^2_{X_{i_0}}\geq c/2$
when $s_k\leq t\leq s_k+\delta$. Without loss of generality, we assume $[s_k,s_k+\delta]\cap\Delta=\emptyset$ for all $k$.
We have
\begin{align}
&\frac{dh_{i_0}(t)}{dt}\leq \sum_{j\in\mathcal{N}_{i_0}(t)}|x_{i_0}(t)|_{X_0}(|x_{j}(t)|_{X_0}-|x_{i_0}(t)|_{X_0})
-\alpha_{i_0,t}|x_{i_0}(t)|^2_{X_{i_0}}+\alpha_{i_0,t}\tan \theta_{i_0,t}|x_{i_0}(t)|^2_{X_0}\nonumber
\end{align}
and then for $s_k\leq t\leq s_k+\delta$,
\begin{align}
\label{aa3}D^+|x_{i_0}(t)|_{X_0}&\leq \sum_{j\in\mathcal{N}_{i_0}(t)}(|x_{j}(t)|_{X_0}-|x_{i_0}(t)|_{X_0})-\frac{\alpha_{i_0,t}|x_{i_0}(t)|^2_{X_{i_0}}}{\phi}
+\alpha_{i_0,t}\tan \theta^+_{t}\phi\\
&\;\leq (n-1)\big(\phi-|x_{i_0}(t)|_{X_0}\big)-
\frac{c}{2\phi}+\alpha^+_{t}\tan \theta^+_{t}\phi,\nonumber
\end{align}
which leads to
\begin{align}
|x_{i_0}(t)|_{X_0}&\leq e^{-(n-1)(t-s_k)}|x_{i_0}(s_k)|_{X_0}+(1-e^{-(n-1)(t-s_k)})\big(\phi-\frac{c}{2(n-1)\phi}\big)\nonumber\\
&+\phi\int^t_{s_k}e^{-(n-1)(t-s)}\alpha^+_s\tan \theta^+_sds\nonumber
\end{align}
and then
\begin{align}\label{number}
&|x_{i_0}(s_k+\delta)|_{X_0}\leq \zeta(\sqrt{2h^*}+\varepsilon)
+(1-\zeta)\Big(\sqrt{2h^*}+\varepsilon-\frac{c}{2(n-1)\phi}\Big)
+\varepsilon\frac{\sqrt{2h^*}+\varepsilon}{\sqrt{2h^*}},
\end{align}
where $0<\zeta=e^{-(n-1)\delta}<1$. We can find that the right-hand side of (\ref{number}) is
less than $\sqrt{2h^*}-\frac{c(1-\zeta)}{4(n-1)\sqrt{2h^*}}$ when $\varepsilon$ is sufficiently small,
which contradicts $\lim_{t\rightarrow\infty}h_{i_0}(t)=h^*$.
Thus, $\lim_{t\rightarrow\infty}\alpha_{i,t}|x_i(t)|^2_{X_i}$ $=0$, $\forall i$. From Theorem \ref{thnew} we have $0\leq \alpha_{i,t}\leq C_i\alpha^*$, and hence $\lim_{t\rightarrow\infty}\alpha_{i,t}|x_{i}(t)|_{X_{i}}=0$, $\forall i$.
According to the equality in (\ref{relation}) and Lemma \ref{lem5}, consensus is achieved for system (\ref{system}).

Step (ii). Suppose $h^+_{i}=h^-_{i}=h^*>0,\forall i$. We will show that $\liminf_{t\rightarrow\infty}\sum^n_{i=1}|x_i(t)|^2_{X_i}=0$
by contradiction.

Hence suppose there is $b>0$ such that $\sum^n_{i=1}|x_i(t)|^2_{X_i}\geq b$ for all sufficiently large $t$.
Let $|x(t)|_{X_0}=(|x_1(t)|_{X_0},...,$ $|x_n(t)|_{X_0})^T$,
$y(t)=(|x_1(t)|^2_{X_1},...,|x_n(t)|^2_{X_n})^T$,\\
$D(t)=\mbox{diag}\{\alpha_{1,t},...,\alpha_{n,t}\}$ (a diagonal matrix with diagonal entries $\alpha_{i,t}$). Then by (\ref{aa3}) we have
\begin{align}\label{numb}
D^+|x(t)|_{X_0}\leq -\mathcal{L}_{\sigma(t)}|x(t)|_{X_0}-\frac{1}{\phi}D(t)y(t)+\phi\alpha^+_t\tan \theta^+_{t}\textbf{1},
\end{align}
where $\mathcal{L}_{\sigma(t)}$ is the Laplacian of graph $\mathcal{G}_{\sigma(t)}$ with
$(\mathcal{L}_{\sigma(t)})_{ij}=-1$ if $j\in \mathcal{N}_i(t)$,
$(\mathcal{L}_{\sigma(t)})_{ij}=0$ if $j\neq i,j\not\in \mathcal{N}_i(t)$ and
$(\mathcal{L}_{\sigma(t)})_{ii}=|\mathcal{N}_i(t)|$.
Recall that $t_k,k\geq0$ are all the switching moments of switching graph $\mathcal{G}_{\sigma}$
with $t_{k+1}-t_k\geq \tau,\forall k$.
It is easy to see that we can add some new ``switching moments'' in $\{t_k\}_{k\geq0}$, denoted as
$\{t'_k\}_{k\geq0}$,
such that  $2\tau\geq t'_{k+1}-t'_k\geq \tau,\forall k$.
From (\ref{numb}) we have
\begin{align}
|x(t'_{k+1})|_{X_0}&\leq e^{-\mathcal{L}_{\sigma(t'_k)}(t'_{k+1}-t'_k)}|x(t'_{k})|_{X_0}+\nonumber\\
&\int^{t'_{k+1}}_{t'_k}e^{-\mathcal{L}_{\sigma(t'_k)}(t'_{k+1}-t)}\big(-\frac{D(t)y(t)}{\phi}
+\phi\alpha^+_t\tan \theta^+_{t}\textbf{1}\big)dt.\nonumber
\end{align}
Note that for any $s>0$, $e^{-s\mathcal{L}_{\sigma(t'_k)}}$ is a stochastic matrix
(with nonnegative entries and all row sums are ones)
and the graph $\mathcal{G}_{\sigma(t'_k)}$ is a subgraph of the graph associated with matrix $e^{-s\mathcal{L}_{\sigma(t'_k)}}$.
Then applying the similar arguments given in the proof of Theorem 4.1 in \cite{lou12}
we can show that $\liminf_{t\rightarrow\infty}\sum^n_{i=1}|x_i(t)|^2_{X_i}=0$.

Then there is a subsequence $\{s_k\}_{k\geq0}$ with $\lim_{k\rightarrow\infty}s_k=\infty$
such that $\lim_{k\rightarrow\infty}|x_i(s_k)|_{X_i}=0$ for all $i$.
Because we have shown that consensus is achieved in Step (i),
$\lim_{k\rightarrow\infty}|x_i(s_k)|_{X_j}=0$ for all $i,j$, which
leads to $\lim_{k\rightarrow\infty}h_i(s_k)=0$ for all $i$.
Thus, $h^*=\lim_{t\rightarrow\infty}h_i(t)=0$,
which contradicts $h^+_{i}=h^-_{i}=h^*>0$.
It follows that $h^+_{i}=h^-_{i}=h^*=0$ and then the first conclusion is proved.

Notice that $\alpha_{i,t}=\alpha_{j,t}=\alpha_t$ $\forall i,j,t$ and $P^h_{X_i}(x_i(t))=P^a_{X_i}(x_i(t))=P_{X_i}(x_i(t))$ in the case of $\theta_{i,t}=0$ $\forall i,t$.
Let $\varpi_i(t)=\alpha_{t}\big(P_{X_i}(x_i(t))-x_i(t)\big)$.
Then (\ref{system}) can be written as
$$\dot x(t)=-(\mathcal{L}_{\sigma(t)}\otimes I_m)x(t)+\varpi(t),$$
where $x(t)=(x^T_1(t),...,x^T_n(t))^T$ and $\varpi(t)=(\varpi^T_1(t),...,\varpi^T_n(t))^T$.
Hence,
$$
x(t'_{k+1})=e^{-(\mathcal{L}_{\sigma(t'_k)}\otimes I_m)(t'_{k+1}-t'_k)}x(t'_k)+\varphi(t'_k),
$$
where $\varphi(t'_k)=\int^{t'_{k+1}}_{t'_k}e^{-(\mathcal{L}_{\sigma(t'_k)}\otimes I_m)(t'_{k+1}-t)}\varpi(t)dt$.
Clearly, $|\varpi_i(t)|\leq\sqrt{2\bar h}\alpha_t$,
and then $|\varphi(t'_k)|\leq\sqrt{2n\bar h}\int^{t'_{k+1}}_{t'_k}\alpha_tdt$.
Similar to the arguments given in Theorem 4.2 in \cite{lou12}, we can show the second conclusion.
\hfill$\square$

\section{Empty Intersection Case}

In this section, we discuss the convergence in the empty intersection case (i.e., $\cap^n_{i=1}X_i=\emptyset$) and then show some properties of the optimal solution set in the following two subsections.

\subsection{Convergence Analysis}
The following is the convergence result for the case when $\cap^n_{i=1}X_i=\emptyset$.

\begin{them}
\label{thm1} Suppose {\bf A1}-{\bf A3} hold,
$\mathcal{G}_{\sigma(t)},t\geq 0$ are undirected and $\cap^n_{i=1}X_i=\emptyset$.
Then SDOP is solved by system (\ref{system}) if
 $\int^{\infty}_{0}\alpha_tdt=\infty$, $\int^{\infty}_{0}\alpha^2_tdt<\infty$
and $\int^{\infty}_{0}\alpha_t\tan\theta^+_tdt<\infty$;  Furthermore,
if $\theta_{i,t}=0\;\forall i,t$,
then it is necessary that $\lim_{t\rightarrow\infty}\alpha_t=0$
for (\ref{system}) to solve SDOP.
\end{them}

\begin{rem}
In the case of the exact projection (i.e., $\theta_{i,t}\equiv0$),
the stepsize conditions in Theorem \ref{thm1} become $\int^{\infty}_{0}\alpha_tdt=\infty$ and
$\int^{\infty}_{0}\alpha^2_tdt<\infty$, which is a continuous-time version of the
discrete-time stochastic approximation stepsize condition $\sum^\infty_{k=0}\alpha_k=\infty$ and $\sum^\infty_{k=0}\alpha^2_k<\infty$ given in \cite{nedic2}
to solve the optimization problem $\min\sum^n_{i=1}f_i$. Therefore, the result in Theorem \ref{thm1}
is consistent with those in \cite{nedic2}.
Note that \cite{lin} proposed distributed continuous-time algorithms
for the empty intersection case when the graphs kept connected, which is more restrictive than the UJSC given in this paper.
\end{rem}

From Theorems \ref{thm2} and \ref{thm1}, we find that
the sufficient optimal consensus conditions are
essentially different in these two cases. In addition to the conditions in the nonempty intersection case,
the square integrability condition is usually required in the empty intersection case.

Before presenting the proof of Theorem \ref{thm1},
we show two lemmas. The first one is taken from Lemma 7 in \cite{nedic2}.

\begin{lem}
\label{lem5.1} Let $0<\lambda<1$ and $\{b_k\}_{k\geq1}$ be a positive sequence.
If $\sum^\infty_{k=1}b_k<\infty$, then $\sum^\infty_{k=1}\sum^k_{r=1}\lambda^{k-r}b_r<\infty$.
\end{lem}

\begin{lem}
\label{lemcon2}
Under {\bf A1}, if $\int^{\infty}_{0}\alpha^2_tdt<\infty$ and $\int^{\infty}_{0}\alpha_t\tan\theta^+_tdt<\infty$, then
\begin{align}
\int^\infty_0 \alpha_t|x_i(t)-\bar x(t)|dt<\infty\nonumber
\end{align}
for all $i$, where $\bar x(t)=\frac{1}{n}\sum^n_{i=1}x_i(t)$.
\end{lem}

\emph{Proof.} Let $H(t)=\max_{1\leq i,j\leq n}|x_i(t)-x_j(t)|$.
Since $|x_i(t)-\bar x(t)|\leq H(t)$,
it suffices to show $\int^\infty_0\alpha_tH(t)dt<\infty$.
By Theorem \ref{bounded} (i),
(\ref{algo0}) and Lemma \ref{lemcon},
for any $k\geq 0$ and $t$, $kB_0\leq t<(k+1)B_0$,
it holds that
\begin{align}
\label{a52}
H((k+1)B_0)&\leq \beta H(kB_0)+B_2\nu(kB_0),\\
\label{a53}
H(t)&\leq H(kB_0)+B_2\nu(kB_0),
\end{align}
for some $B_2>0$, where $\nu(kB_0)=\int^{(k+1)B_0}_{kB_0}\alpha_tdt$.
Thus, with (\ref{a53}) we have
\begin{align}\label{a54}
\int^\infty_0 \alpha_tH(t)dt&=\sum^\infty_{k=0}\int^{(k+1)B_0}_{kB_0}\alpha_tH(t)dt\nonumber\\
&\leq\sum^\infty_{k=0}\int^{(k+1)B_0}_{kB_0}\alpha_t\big(H(kB_0)+B_2\nu(kB_0)\big)dt\nonumber\\
&=\sum^\infty_{k=0}\nu(kB_0)\big(H(kB_0)+B_2\nu(kB_0)\big).
\end{align}

Now we estimate the term in (\ref{a54}).
First, by Cauchy-Schwarz inequality $\int g_1g_2\leq\sqrt{\int g^2_1\int g^2_2}$, we have
\begin{align}\label{a55}
\sum^\infty_{k=0}\nu^2(kB_0)
\leq B_0\sum^\infty_{k=0}\int^{(k+1)B_0}_{kB_0}\alpha^2_tdt
=B_0\int^{\infty}_0\alpha^2_tdt
<\infty.
\end{align}
Second, by (\ref{a52}) we have
$H(kB_0)\leq \beta^k H(0)+B_2\sum^k_{r=1}\beta^{k-r}\nu((r-1)B_0),\forall k\geq1$.
Thus,
\begin{align}\label{a56}
\sum^\infty_{k=1}H(kB_0)\nu(kB_0)
&\leq  H(0)\sum^\infty_{k=1}\beta^k\nu(kB_0)
+B_2\sum^\infty_{k=1}\nu(kB_0)\sum^k_{r=1}\beta^{k-r}\nu((r-1)B_0)\nonumber\\
&\leq \frac{H(0)\beta\alpha^*B_0}{1-\beta}
+\frac{B_2}{2}\sum^\infty_{k=1}\sum^k_{r=1}\beta^{k-r}\big(\nu^2(kB_0)+\nu^2((r-1)B_0)\big)\nonumber\\
&\leq \frac{H(0)\beta\alpha^*B_0}{1-\beta}
+\frac{B_2}{2(1-\beta)}\sum^\infty_{k=1}\nu^2(kB_0)+\frac{B_2}{2}\sum^\infty_{k=1}\sum^k_{r=1}\beta^{k-r}
\nu^2((r-1)B_0)\nonumber\\
&<\infty,
\end{align}
where the second inequality follows from $\nu(kB_0)\leq \alpha^*B_0$, 
the third one from $\sum^k_{r=1}\beta^{k-r}\leq \frac{1}{1-\beta},\forall k$,
and the last one from (\ref{a55}) and Lemma \ref{lem5.1}.
Thus, the conclusion follows from (\ref{a54}), (\ref{a55}) and (\ref{a56}). \hfill$\square$

It is time to give the proof of Theorem \ref{thm1}.

\vskip 2mm

\emph{Proof of Theorem \ref{thm1}.}
We rewrite (\ref{system}) as
$$
\dot x_i(t)=\sum_{j\in\mathcal{N}_i(t)}(x_j(t)-x_i(t))+\alpha_{t}(P^h_{X_i}(x_i(t))-x_i(t))+\phi_i(t),
$$
where $\phi_i(t)=(\alpha_{i,t}-\alpha_{t})(P^h_{X_i}(x_i(t))-x_i(t))$.
From the definition of $P^h_{X_i}$, we have
$|P^h_{X_i}(x_i(t))-x_i(t)|=\frac{|x_i(t)|_{X_i}}{\cos\theta_{i,t}}\leq\frac{\eta}{\cos\theta_{i,t}},$
where
$$\eta=\sup_{i,j,t}\{|x_i(t)-x^*|,|\bar x(t)|_{X_i},|x_i(t)|_{X_j}\}$$
is a finite number by Theorem \ref{bounded}.
Moreover, it follows from (\ref{rg}) that $|\alpha_{i,t}-\alpha_{t}|\leq (C_{i,t}-1)\alpha_{t}\leq\frac{1}{\sin\mu_{i}}\alpha_{t}\sin \theta_{i,t}$. Then
from the preceding two estimates,
\begin{align}\label{err}
|\phi_i(t)|\leq \frac{\eta}{\sin\mu_{i}}\alpha_{t}\tan\theta_{i,t}.
\end{align}

Take $x^*\in \arg\min \sum^n_{i=1}|x|^2_{X_i}$. Let $t\not\in \Delta$.
Clearly,
\begin{align}
\frac{d|x_i(t)-x^*|^2}{dt}&=2\big\langle x_i(t)-x^*,\dot x_i(t)\big\rangle\nonumber\\
&=2\Big\langle x_i(t)-x^*,\sum_{j\in\mathcal{N}_{i}(t)}(x_j(t)-x_i(t))+\alpha_t(P^h_{X_i}(x_i(t))-x_i(t))+\phi_i(t)\Big\rangle.\nonumber
\end{align}
Because $\mathcal{G}_{\sigma(t)}$ is undirected,
\begin{align}\label{a48}
&\frac{d\sum^n_{i=1}|x_i(t)-x^*|^2}{dt}\nonumber\\
&=-2\sum_{j\in\mathcal{N}_{i}(t)}\big|x_j(t)- x_i(t)\big|^2+2\sum^n_{i=1}\big\langle x_i(t)-x^*,\alpha_t (P^h_{X_i}(x_i(t))-x_i(t))+\phi_i(t)\big\rangle\nonumber\\
&\leq2\alpha_t\sum^n_{i=1}\big\langle x_i(t)-x^*,P^h_{X_i}(x_i(t))-x_i(t)\big\rangle
+2\sum^n_{i=1}\big\langle x_i(t)-x^*,\phi_i(t)\big\rangle.
\end{align}

Then we estimate the first term in (\ref{a48}).  Note that
\begin{align}\label{a9}
\big\langle x_i(t)-x^*,P^h_{X_i}(x_i(t))-P_{X_i}(x_i(t))\big\rangle
\leq |x_i(t)-x^*|\tan \theta_{i,t}|x_i(t)|_{X_i}\leq \eta^2\tan \theta^+_t.
\end{align}
 We also have
\begin{align}\label{a49}
\sum^n_{i=1}\big\langle x_i(t)-x^*,P_{X_i}(x_i(t))-x_i(t)\big\rangle
&=-\big\langle \bar x(t)-x^*,\sum^n_{i=1}\big(\bar x(t)-P_{X_i}(\bar x(t))\big)\big\rangle+\varrho(t),
\end{align}
where
$$\varrho(t)=\sum^n_{i=1}\big\langle x_i(t)-\bar x (t),P_{X_i}(\bar x(t))-\bar x(t)\big\rangle
+\sum^n_{i=1}\big\langle x_i(t)-x^*,P_{X_i}(x_i(t))-P_{X_i}(\bar x(t))+\bar x(t)-x_i(t)\big\rangle.$$
Clearly, the first term in $\varrho(t)$ is not greater than $\eta\sum^n_{i=1}|x_i(t)-\bar x(t)|$
and the second term in $\varrho(t)$ is not greater than $2\eta\sum^n_{i=1}|x_i(t)-\bar x(t)|$ by Lemma \ref{pro} (iv).
Moreover, by (\ref{err}), the second term in (\ref{a48}) is not greater than
$2n\eta^2\alpha_{t}\tan\theta^+_{t}/\sin\mu_-$, where $\mu_-=\min_{1\leq i\leq n}\mu_i$.
Denote $\psi(t)=\langle\bar x(t)-x^*,\sum^n_{i=1}(\bar x(t)-P_{X_i}(\bar x(t)))\rangle$,
$\varsigma(t)=6\eta(\sum^n_{i=1}\alpha_t|x_i(t)-\bar x(t)|)
+2n\eta^2(1+\frac{1}{\sin\mu_-})\alpha_t\tan \theta^+_{t}.$
In light of (\ref{a48}), (\ref{a9}) and (\ref{a49}),  we have
\begin{align}\label{a50}
\frac{d\sum^n_{i=1}|x_i(t)-x^*|^2}{dt}&\leq -2\alpha_t\psi(t)+\varsigma(t)\leq\varsigma(t)
\end{align}
because $\psi(t)$ is nonnegative, following from
(\ref{conve}) and the convexity of objective function $f$, that is,
\begin{align}\label{add}
\psi(t)=\big\langle\bar x(t)-x^*,\frac{1}{2}\nabla f(\bar x(t))\big\rangle\geq\frac{1}{2}(f(\bar x(t))-f(x^*))\geq0.
\end{align}

We next show that $\lim_{t\to\infty}\sum^n_{i=1}|x_i(t)-x^*|^2$ is a finite number by contradiction.
Let us suppose there exist $\{s_k\}_{k\geq0}$ with $s_k\rightarrow\infty$ and $\varepsilon>0$ such that
$\sum^n_{i=1}|x_i(s_{2k+1})-x^*|^2-\sum^n_{i=1}|x_i(s_{2k})-x^*|^2\geq \varepsilon$ for all $k$.
According to Lemma \ref{lemcon2}, $\int^\infty_0\sum^n_{i=1}\alpha_t|x_i(t)-\bar x(t)|dt<\infty.$
Therefore, there is $K_0>0$ such that
$\int^\infty_{K_0}\sum^n_{i=1}\alpha_t|x_i(t)-\bar x(t)|dt\leq \frac{\varepsilon}{24\eta}$
and $\int^\infty_{K_0}\alpha_{t}\tan \theta^+_{t} dt\leq \frac{\varepsilon}{8n\eta^2(1+1/\sin\mu_-)}$.
By (\ref{a50}), we have that, for sufficiently large $k$ for which $t_{2k}\geq K_0$,
\begin{align}
\varepsilon&\leq\sum^n_{i=1}|x_i(s_{2k+1})-x^*|^2-\sum^n_{i=1}|x_i(s_{2k})-x^*|^2\nonumber\\
&\leq 6\eta\int^\infty_{K_0}\alpha_t\big(\sum^n_{i=1}|x_i(t)-\bar x(t)|\big)dt
+2n\eta^2\int^\infty_{K_0}\alpha_{t}\tan \theta^+_{t}dt\nonumber\\
&\leq \frac{\varepsilon}{2},\nonumber
\end{align}
which yields a contradiction. Hence, $\lim_{t\to\infty}\sum^n_{i=1}|x_i(t)-x^*|^2$ is a finite number.

Thus, it follows from (\ref{a50}) that
$$
2\int^\infty_{0}\alpha_t\big\langle \bar x(t)-x^*,
\sum^n_{i=1}\big(\bar x(t)-P_{X_i}(\bar x(t))\big)\big\rangle=
\int^\infty_{0}\alpha_t\big\langle \bar x(t)-x^*,
\nabla f(\bar x(t))\big\rangle<\infty.
$$
Due to $\int^\infty_{0}\alpha_tdt=\infty$, there is a subsequence
$\{t_r\}_{r\geq0}$ such that $\lim_{r\to\infty}\big\langle \bar x(t_r)-x^*,
\nabla f(\bar x(t_r))\big\rangle=0$.
Since the system states are bounded, without loss of generality we assume
$\lim_{r\to\infty}\bar x(t_r)=\hat x$ for some $\hat x$ (otherwise we can further find a subsequence
of $\{t_r\}_{r\geq0}$). Since $\nabla f$ is continuous,
$\big\langle \hat x-x^*, \nabla f(\hat x)\big\rangle=0$, which
leads to $f(x^*)\geq f(\hat x)+\langle x^*-\hat x,\nabla f(\hat x)\rangle=f(\hat x)$.
Thus, $\hat x\in \arg\min f$.

Replacing $x^*$ with $\hat x$,
we can similarly show that $\lim_{t\to\infty}\sum^n_{i=1}|x_i(t)-\hat x|^2$ is also a finite number, denoted as $\rho$.
Moreover, the uniform continuity of $\alpha_t$ and $\int^\infty_{0}\alpha^2_tdt<\infty$
imply $\lim_{t\to\infty}\alpha_{t}=0$. Therefore, consensus is achieved by Lemma \ref{lem5}
and then $\lim_{r\to\infty}x_i(t_r)=\hat x$.
Hence $\rho=0$, which in return implies $\lim_{t\to\infty}x_i(t)=\hat x$ for all $i$. Then the first part is completed.

Now we show the second part. Notice that $\theta_{i,t}\equiv0$
implies that $\alpha_{i,t}=\alpha_{j,t}=\alpha_t\;\forall i,j,t$
and $P^h_{X_i}(x_i(t))=P^a_{X_i}(x_i(t))=P_{X_i}(x_i(t))$.
Let $x^*\in\arg\min\sum^n_{i=1}|x|^2_{X_i}$ be the point with
$\lim_{t\rightarrow\infty}x_i(t)=x^*$ for all $i$.
According to the boundedness of system states, the non-expansive property Lemma \ref{pro} (iv) and the uniform continuity of
$\alpha_t$, $\dot x_i(t)$ is uniformly continuous with respect to time intervals $(t_k,t_{k+1})$, $k\geq0$, which, along with $\lim_{t\rightarrow\infty}x_i(t)=x^*$, leads to
$\lim_{t\rightarrow\infty}\dot x_i(t)=0$  by Lemma \ref{mod}.
Therefore, because the network achieves a consensus,
$$\lim_{t\rightarrow\infty}\alpha_t(P_{X_i}(x_i(t))-x_i(t))
=\lim_{t\rightarrow\infty}\alpha_t(P_{X_i}(x^*)-x^*)=0,i=1,...,n.$$
According to $\cap^n_{i=1}X_i=\emptyset$, $x^*\not\in X_i$
for at least one $i$.  Thus, $\lim_{t\rightarrow\infty}\alpha_t=0$. \hfill$\square$
%

\subsection{Optimal Solutions}

Theorem \ref{thm1} showed that all agents consensually converge to
an optimal solution of $\min\sum^n_{i=1}|x|^2_{X_i}$ under certain conditions.
Next, we show some properties of the optimal solution set of $\min\sum^n_{i=1}|x|^2_{X_i}$, denoted as $X^*$.
According to Lemma \ref{aub}, the optimal solution $x^*\in X^*$ 
must satisfy $$\nabla\sum^n_{i=1}|x^*|^2_{X_i}=2\sum^n_{i=1}(x^*-P_{X_i}(x^*))=0,$$
 or equivalently,
$x^*=\frac{\sum^n_{i=1}P_{X_i}(x^*)}{n}.$  Then we have the following results.

Before showing some properties of the optimal solutions, we give a lemma first.

\begin{lem}\label{pro1}
Let $K$ be a closed convex set in $\mathbb{R}^m$. Then
\begin{align}
(i)\quad&\langle y-z,P_K(y)-P_K(z)\rangle\geq|P_K(y)-P_K(z)|^2\;\;\mbox{for any}\;\; y\;\mbox{and}\;z;\nonumber\\
(ii)\quad&|P_K(y)-P_K(z)|=|y-z|\;\;\mbox{if and only if}\;y-P_K(y)=z-P_K(z).\nonumber
\end{align}
\end{lem}

\emph{Proof.} (i) follows from
\begin{align}
\langle y-z,P_K(y)-P_K(z)\rangle&=\langle y-P_K(y),P_K(y)-P_K(z)\rangle+\big|P_K(y)-P_K(z)\big|^2\nonumber\\
&\qquad\qquad\qquad\qquad\quad+\langle P_K(z)-z,P_K(y)-P_K(z)\rangle\nonumber\\
&\geq\big|P_K(y)-P_K(z)\big|^2\nonumber
\end{align}
because $\langle y-P_K(y),P_K(y)-P_K(z)\rangle\geq0$
and $\langle P_K(z)-z,P_K(y)-P_K(z)\rangle\geq0$ by Lemma \ref{pro} (i).

For (ii), the sufficiency is obvious. The necessity can be obtained from
\begin{align}
\big|y-P_K(y)-(z-P_K(z))\big|^2&=|y-z|^2+\big|P_K(z)-P_K(y)\big|^2+2\langle y-z,P_K(z)-P_K(y)\rangle\nonumber\\
&=2|y-z|^2+2\langle y-z,P_K(z)-P_K(y)\rangle\nonumber\\
&\leq2|y-z|^2-2\big|P_K(y)-P_K(z)\big|^2\nonumber\\
&=0,\nonumber
\end{align}
where the inequality follows from (i) of this lemma.
\hfill$\square$

Let $X^*$ be the optimal solution set of $\min\sum^n_{i=1}|x|^2_{X_i}$.   Then we have the following results.

\begin{them}
(i) For any $x^*$, $y^*\in X^*$, we have $x^*-P_{X_i}(x^*)=y^*-P_{X_i}(y^*),\;\;i=1,...,n;$

(ii) For any $i$, either $X^*\subseteq X_i$ or $X^*\cap X_i=\emptyset$;

(iii)
Let $x^*\in X^*$, $x^*\not\in X_i$ for some $i$.
Then $X^*\cap$line$(x^*,P_{X_i}(x^*))=\{x^*\}$.
\end{them}

\emph{Proof.} (i) Since $x^*=\frac{\sum^n_{i=1}P_{X_i}(x^*)}{n}$
and $y^*=\frac{\sum^n_{i=1}P_{X_i}(y^*)}{n}$,
\begin{align}
|x^*-y^*|&=\Big|\frac{\sum^n_{i=1}(P_{X_i}(x^*)-P_{X_i}(y^*))}{n}\Big|\nonumber\\
&\leq\frac{\sum^n_{i=1}|P_{X_i}(x^*)-P_{X_i}(y^*)|}{n}\nonumber\\
&\leq |x^*-y^*|\nonumber
\end{align}
from Lemma \ref{pro} (iv). Therefore, $|P_{X_i}(x^*)-P_{X_i}(y^*)|=|x^*-y^*|$ for all $i$, which implies
the conclusion by Lemma \ref{pro1} (ii).

(ii) This is straightforward from (i).

(iii) Let $z^*\in X^*\cap$ line$(x^*,P_{X_i}(x^*))$, $z^*\neq x^*$.
If $z^*$ locates the half-line with $P_{X_i}(x^*)$ as the starting point
and $x^*-P_{X_i}(x^*)$ as the direction,
then $P_{X_i}(z^*)=P_{X_i}(x^*)$ by Lemma \ref{pro} (i) (note that Lemma \ref{pro} is an equivalent definition of convex projection).
Therefore, $x^*-P_{X_i}(x^*)\not=z^*-P_{X_i}(z^*)$, which contradicts
what we have proven in (i) since both $x^*$ and $z^*$ are optimal solutions.
If $z^*$ locates the half-line with $P_{X_i}(x^*)$ as the starting point
and $P_{X_i}(x^*)-x^*$ as the direction, then $P_{X_i}(x^*)$
is also an optimal solution since the optimal solution set $X^*$
is a convex set and $P_{X_i}(x^*)$ can be written as a convex combination of
$x^*,z^*$. Then $0=P_{X_i}(x^*)-P_{X_i}(P_{X_i}(x^*))\neq x^*-P_{X_i}(x^*)$,
which also yields a contradiction since both $x^*$ and $P_{X_i}(x^*)$ are optimal solutions. Thus, the conclusion follows.
\hfill$\square$

\section{Fixed Stepsize and Approximate Angle}

In this section, we consider the constant stepsize and approximate angle case.
The following result is about the convergence error between the agents' estimates and the optimal point
in terms of the stepsize and approximate angle.

\begin{them}\label{constant}
Consider system (\ref{algo0}) with $\alpha_{t}\equiv\alpha>0$ and $0\leq\theta_{i,t}\equiv\theta_i<\pi/2$, under {\bf A1}-{\bf A3} and that $\mathcal{G}_{\sigma(t)}$ is undirected for $t\geq0$.

(i) Suppose $\cap^n_{i=1}int(X_i)\neq\emptyset$ (implies $\cap^n_{i=1}X_i\neq\emptyset$). Then
\begin{align}
\limsup_{t\rightarrow\infty}|x_i(t)|_{X_0}\leq
&\sqrt{\kappa_{S}n\eta d_0\Big(\frac{4-2\beta}{1-\beta}
B_0B_1C\sqrt{1+(\tan\theta^+)^2}\alpha+C\tan\theta^+\Big)}\nonumber\\
&+\frac{2-\beta}{1-\beta}B_0B_1C\sqrt{1+(\tan\theta^+)^2}\eta\alpha.\nonumber
\end{align}

(ii) Suppose $\cap^n_{i=1}X_i=\emptyset$ and
$f(x)=\sum^n_{i=1}|x|^2_{X_i}$ is $\ell$-strongly convex.
Let $x^*$ be the unique optimal solution of $\min f$.
Then
\begin{align}
\limsup_{t\rightarrow\infty}|x_i(t)-x^*|\leq&\sqrt{\frac{4 n\eta^2}{\ell}\Big(\frac{4-2\beta}{1-\beta}B_0B_1C\sqrt{1+(\tan\theta^+)^2}\alpha+C\tan\theta^+\Big)}\nonumber\\
&+\frac{2-\beta}{1-\beta}B_0B_1C\sqrt{1+(\tan\theta^+)^2}\eta\alpha\nonumber
\end{align}
with $S=X_c+\mathbf{B}(0,r_0)$, $C=\max_{1\leq i\leq n}C_i$,
$\theta^+=\max_{1\leq i\leq n}\theta_i$, $C_i$ defined in (\ref{rg}),
$\kappa_{S}$ defined in Lemma \ref{dis}, and
$\beta$, $B_0$ and $B_1$ defined in Lemma \ref{lemcon}, where
$d_0=\sup_{i,t}|x_i(t)|_{X_0}$, which is finite by Theorem \ref{bounded}.
\end{them}


\emph{Proof.} Similar to (\ref{a52}),
$H((k+1)B_0)\leq \beta H(kB_0)+B_0B_1C\sqrt{1+(\tan\theta^+)^2}\eta\alpha$,
and then
$$H(kB_0)\leq \beta^k H(0)+(1+\beta+\cdots+\beta^{k-1})B_0B_1C\sqrt{1+(\tan\theta^+)^2}\eta\alpha,$$
along with $H(t)\leq H(kB_0)+B_0B_1C\sqrt{1+(\tan\theta^+)^2}\eta\alpha,\forall kB_0\leq t<(k+1)B_0$, implies
\begin{align}
H(t)&\leq \beta^k H(0)+(2+\beta+\cdots+\beta^{k-1})B_0B_1C\sqrt{1+(\tan\theta^+)^2}\eta\alpha\nonumber\\
&\leq\frac{H(0)}{\beta^{\frac{1}{B_0}}}(\beta^{\frac{1}{B_0}})^t
+\frac{2-\beta}{1-\beta}B_0B_1C\sqrt{1+(\tan\theta^+)^2}\eta\alpha\nonumber
\end{align}
when $kB_0\leq t<(k+1)B_0$.   Then
\begin{align}\label{ddd}
\limsup_{t\rightarrow\infty}H(t)\leq\frac{2-\beta}{1-\beta}B_0B_1C\sqrt{1+(\tan\theta^+)^2}\eta\alpha.
\end{align}
Clearly, (\ref{algo0}) can be written as
\begin{align}\label{sy}
\dot x_i(t)=\sum_{j\in\mathcal{N}_i(t)}(x_j(t)-x_i(t))+\alpha(P_{X_i}(\bar x(t))-\bar x(t))+\omega_i(t),\;i=1,...,n,
\end{align}
where $\bar x(t)=\frac{1}{n}\sum^n_{i=1}x_i(t)$,
$$
\omega_i(t)=\alpha\big(P_{X_i}(x_i(t))-P_{X_i}(\bar x(t))+\bar x(t)-x_i(t)\big)
+\alpha(P^a_{X_i}(x_i(t))-P_{X_i}(x_i(t))).
$$
The first term in $\omega_i(t)$ is not greater than $2\alpha H(t)$
by Lemma \ref{pro} (iv) and the inequality $|\bar x(t)-x_i(t)|\leq H(t)$, and
the second term is not greater than $C\eta\alpha\tan\theta^+$ due to the
relation
$$\frac{|P^a_{X_i}(x_i(t))-P_{X_i}(x_i(t))|}{|P^h_{X_i}(x_i(t))-P_{X_i}(x_i(t))|}
=\frac{\sin(\frac{\pi}{2}+\theta_{i,t})}{\sin\mu_{i,t}}\leq\frac{1}{\sin\mu_{i}}\leq C_i,$$
where the equality follows from the well-known law of sines:
$$\frac{a_1}{\sin A_1}=\frac{a_2}{\sin A_2}=\frac{a_3}{\sin A_3}$$
with $a_1$, $a_2$, $a_3$ the lengths of the sides of a triangle, and $A_1$, $A_2$, $A_3$ the opposite angles. Hence,
\begin{align}\label{sy2}
|\omega_i(t)|\leq 2\alpha H(t)+C\eta\alpha\tan\theta^+.
\end{align}
Therefore, from (\ref{sy}) and the undirectedness of $\mathcal{G}_{\sigma(t)}$, we have
$$
\dot{\bar x}(t)=\frac{1}{n}\sum^n_{i=1}\dot x_i(t)
=\frac{\alpha}{n}\sum^n_{i=1}\big(P_{X_i}(\bar x(t))-\bar x(t)\big)+\frac{1}{n}\sum^n_{i=1}\omega_i(t).
$$

We complete the proof for both the nonempty intersection and empty intersection case.

(i) If $\cap^n_{i=1}int(X_i)\neq\emptyset$, then, from Lemma \ref{aub}, for any $t\not\in\Delta$,
\begin{align}\label{sy4}
\frac{d|\bar x(t)|^2_{X_0}}{dt}
&=\frac{2\alpha}{n}\big\langle\bar x(t)-P_{X_0}(\bar x(t)),\sum^n_{i=1}\big(P_{X_i}(\bar x(t))-\bar x(t)\big)\big\rangle
+\frac{2}{n}\big\langle\bar x(t)-P_{X_0}(\bar x(t)),\sum^n_{i=1}\omega_i(t)\big\rangle\nonumber\\
&\leq-\frac{2\alpha}{n}\sum^n_{i=1}|\bar x(t)|^2_{X_i}+\big(4\alpha H(t)+2C\eta\alpha\tan\theta^+\big)d_0\nonumber\\
&\leq-\frac{2\alpha}{n}\max_{1\leq i\leq n}|\bar x(t)|^2_{X_i}+\big(4\alpha H(t)+2C\eta\alpha\tan\theta^+\big)d_0\nonumber\\
&\leq-\frac{2\alpha}{\kappa_{S} n}|\bar x(t)|^2_{X_0}+\big(4\alpha H(t)+2C\eta\alpha\tan\theta^+\big)d_0,
\end{align}
where
the first inequality follows from (\ref{a23}) (replacing $x_i(t)$ with $\bar x(t)$) and (\ref{sy2});
the third one from Lemma \ref{dis}. As a result, we obtain that for any $t\geq t_0\geq0$,
$$
|\bar x(t)|^2_{X_0}\leq e^{-\frac{2\alpha}{\kappa_{S}n}(t-t_0)}|\bar x(t_0)|^2_{X_0}
+e^{-\frac{2\alpha}{\kappa_{S} n}t}\int^t_{t_0}e^{\frac{2\alpha}{\kappa_{S} n}s}\big(4\alpha H(s)+2C\eta\alpha\tan\theta^+\big)d_0ds,
$$
which combines with (\ref{ddd}) imply
$$\limsup_{t\rightarrow\infty}|\bar x(t)|^2_{X_0}\leq\frac{\kappa_{S} nd_0}{2\alpha}\Big(4\alpha^2\frac{2-\beta}{1-\beta}B_0B_1C\sqrt{1+(\tan\theta^+)^2}\eta+2C\eta\alpha\tan\theta^+\Big).$$
This implies the conclusion by noticing the relation $|x_i(t)|_{X_0}-|\bar x(t)|_{X_0}\leq |\bar x(t)-x_i(t)|\leq H(t)$
and (\ref{ddd}).

(ii) If $\cap^n_{i=1}X_i=\emptyset$ and $f(x)=\sum^n_{i=1}|x|^2_{X_i}$ is $\ell$-strongly convex,
then when $t\not\in\Delta$,
\begin{align}
\frac{d|\bar x(t)-x^*|^2}{dt}
&=\frac{2\alpha}{n}\big\langle\bar x(t)-x^*,\sum^n_{i=1}\big(P_{X_i}(\bar x(t))-\bar x(t)\big)\big\rangle
+\frac{2}{n}\big\langle\bar x(t)-x^*,\sum^n_{i=1}\omega_i(t)\big\rangle\nonumber\\
&\leq-\frac{\alpha}{n}(f(\bar x(t))-f(x^*))+\big(4\alpha H(t)+2C\eta\alpha\tan\theta^+\big)\eta\nonumber\\
&\leq-\frac{\alpha\ell}{2n}|\bar x(t)-x^*|^2+\big(4\alpha H(t)+2C\eta\alpha\tan\theta^+\big)\eta,\nonumber
\end{align}
where the first inequality follows from (\ref{add}) and the second one from (\ref{conve1}).
Thus, the conclusion can be obtained with a proof similar to that for (i). \hfill$\square$


\section{Numerical Examples}

In this section, we provide an example to illustrate
the above convergence results.

Consider a network of three agents with node set $\mathcal{V}=\{1,2,3\}$.
The convex set $X_i$ of each agent $i$ is the ball in $\mathbb{R}^2$
with center $c_i$ and radius $r_i$. Let $\alpha_t=\frac{20}{t+20}$, $\theta_{i,t}=\frac{1}{t+50}$,
which satisfy the conditions in Theorems \ref{thm2} and \ref{thm1}.
We next present the state trajectories of the three agents
for the nonempty and empty intersection cases from time $t=0$ to $t=2000$, respectively.

\begin{figure}[!htbp]
\centering
\includegraphics[width=5.0in]{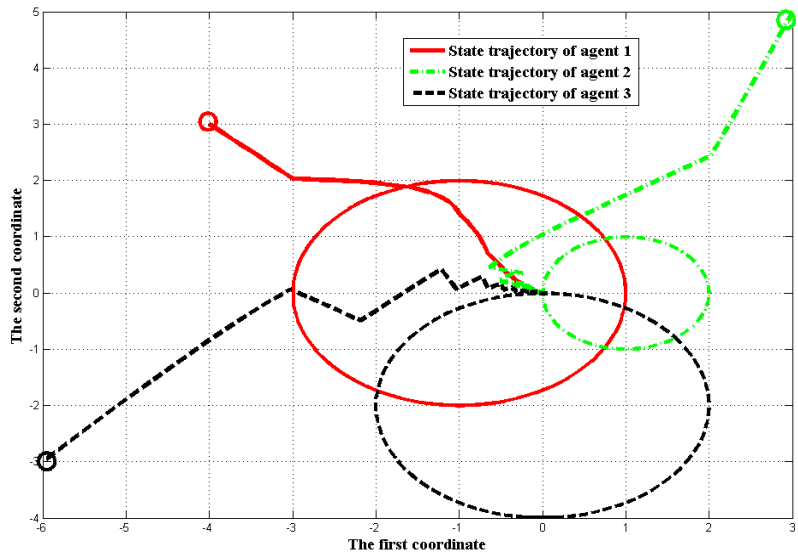}
\caption{In the nonempty intersection case, all agents converge to a common point in the intersection set.}\label{nonempty}
\end{figure}

(i) Nonempty intersection case with
$$c_1=(-1,0)^T,\; c_2=(1,0)^T,\; c_3=(0,-2)^T,\; r_1=2,\; r_2=1, \; r_3=2.$$
The graphs are periodically switching
over the two directed graphs $G_1=(\mathcal{V},\mathcal{E}_1)$,
$G_2=(\mathcal{V},\mathcal{E}_2)$ with period $1$,
where $\mathcal{E}_1=\{(2,1),(3,2)\}$, $\mathcal{E}_2=\{(1,3)\}$.
The initial conditions are $x_1(0)=(-4, 3)$, $x_2(0)=(3, 5)$, $x_3(0)=(-6, -3)$, which are marked as $\circ$
in Fig. \ref{nonempty}.

\begin{figure}[!htbp]
\centering
\includegraphics[width=5.0in]{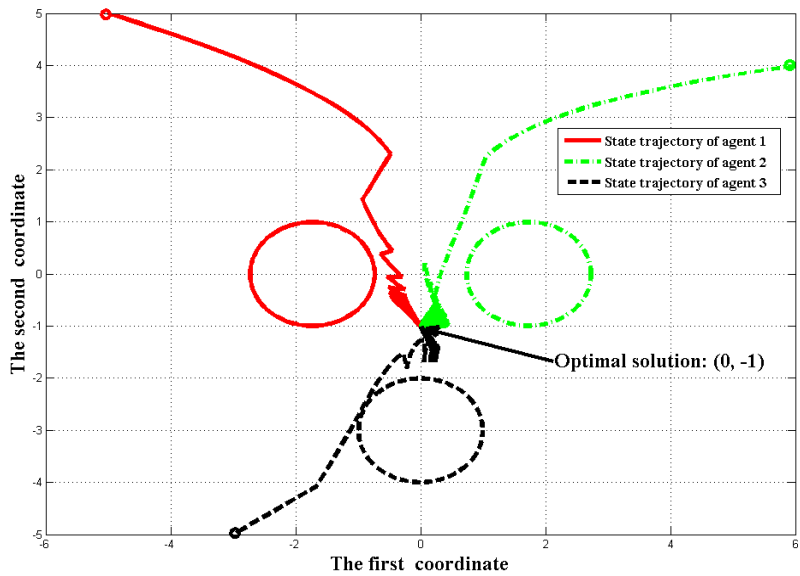}
\caption{In the empty intersection case, all agents converge to the unique optimal solution $(0, -1)$.}\label{empty}
\end{figure}

(ii) Empty intersection case with
$$c_1=(-\sqrt{3},0)^T,\; c_2=(\sqrt{3},0)^T, \; c_3=(0,-3)^T,\;r_1=r_2=r_3=1.$$
In this case, the (unique) optimal solution is $(0,-1)$.
The graphs are periodically switching
over the two undirected graphs $G_1=(\mathcal{V},\mathcal{E}_1)$,
$G_2=(\mathcal{V},\mathcal{E}_2)$ with period $1$,
where $\mathcal{E}_1=\{(3,2)\}$, $\mathcal{E}_2=\{(1,2)\}$.
The initial conditions are $x_1(0)=(-3, 3)$, $x_2(0)=(4, 2)$, $x_3(0)=(-5, -3)$,
which are marked as $\circ$ in Fig. \ref{empty}.

%
%
%

\section{Conclusions}

In this paper, a continuous-time method was proposed to cooperatively solve
the SDOP by a group of agents with the help of graph theory, convex analysis and geometric technique.  Here agents could only obtain
their approximate projections and the communication graph among agents was UJSC.
It was shown that the system states were always bounded for any approximate
angle, and uniformly bounded for any stepsize with inferior limit greater than zero.
Both nonempty intersection and empty intersection cases of convex sets were
investigated with respective sufficient conditions.
Moreover, the convergence error between agents' estimates and the optimal point was also obtained for the constant stepsize and approximate angle case.

\section*{Acknowledgment}

The authors would like to thank Dr. Guilin Yang for
discussions about geometric analysis
and Mr. Peng Yi for his generous help on numerical simulations.

\section*{Appendix}

Denote $\vartheta_i(v)=\angle(v-P_{X_i}(v),P^a_{X_i}(v)-P_{X_i}(v))$.
By Lemma \ref{pro} (i), $\vartheta_i(v)\geq\pi/2$ when $P^a_{X_i}(v)\neq P_{X_i}(v)$.
In the following proofs, we omit all subscript $i$ and
simplify $\theta_i$, $\vartheta_i$, $\mu_i$, $X_i$
as $\theta$, $\vartheta$, $\mu$, $X$.

\textbf{Proof of Lemma \ref{props}.}
Let $v\in S\backslash X$ and $P^a_X(v)\neq P_X(v)$.
We obtain
\begin{align}
\gamma_X(v)&=\frac{|P^h_X(v)-v|+|P^a_X(v)-P^h_X(v)|}{|P^h_X(v)-v|}\nonumber\\
&=1+\frac{|P^a_X(v)-P^h_X(v)|}{|P^h_X(v)-P_X(v)|}\sin \theta(v)\nonumber\\
&=1+\frac{\sin(\vartheta(v)-\frac{\pi}{2})}{\sin\mu(v)}\sin \theta(v)\nonumber\\
&\leq1+\frac{1}{\sin\mu(v)}\sin \theta(v)\label{ef}\\
&\leq1+\frac{1}{\sin\mu(v)}.\nonumber
\end{align}
Then the proof is completed. \hfill$\square$

\begin{figure}[!htbp]
\centering
\includegraphics[width=2.4in]{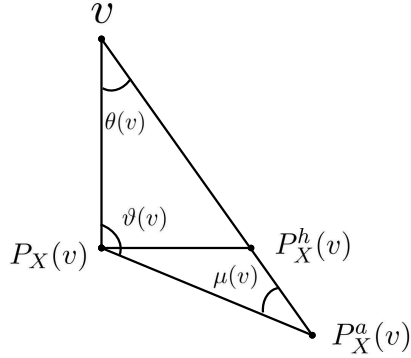}
\caption{An illustration for the proof of Lemma \ref{props}.}\label{ang}
\end{figure}

\vskip 3mm
\textbf{Proof of Lemma \ref{props2}.}
Consider the following relation
\begin{align}\label{co}
cone(v,\mathbf{C}_{X}(v,\theta)\cap \mathbf{b}(v,X))=\mathbf{C}_{X}(v,\theta)
\end{align}
where cone$(v,M)=\{v+\lambda(z-v)|\lambda\geq0,z\in M\}$ for some set $M\subseteq \mathbb{R}^m$. We first show the following three claims:

(i) Suppose that, for any $v\in$ bd$(S)$, there is $\theta(v)>0$ such that
(\ref{co}) holds for $v$, $\theta(v)$.
Then (\ref{co}) holds for $\theta^*$ and any $v\in S\backslash X$ with sufficiently small $|v|_{X}$,
 $\theta^*$ is the approximate angle given in {\bf A3}.

(ii) Suppose int$(X)\cap$line$(v,P_X(v))\neq\emptyset$ for any $v\in$ bd$(S)$. Then
the hypothesis in (i) holds.

(iii) Suppose the boundary surface of $X$ is regular. Then hypothesis in (ii) holds.

(i) is obvious.
For (ii), let $z\in$int$(X)\cap$line$(v,P_X(v))$.
Then there exists $\epsilon>0$ such that $\mathbf{B}(z,\epsilon)\subseteq X$.
Let $y\in$ bd$(\mathbf{B}(z,\epsilon))$ be the point for which
$\angle(y-z,v-z)=\pi/2$.
Clearly, (\ref{co}) holds for $v,\theta(v)$,
where $\theta(v)=\angle(y-v,z-v)>0$.

We prove (iii) by contradiction.
For a regular surface, its tangent plane
at boundary point $z$ consists of the tangent vectors at point $z$ of all curves
passing $z$.  Suppose that there is $v\in$ bd$(S)$ with int$(X)\cap$line$(v,P_X(v))=\emptyset$.
Then, by convex set separation Theorem 11.3 on page 97 in \cite{Roc}, there exists a hyperplane $\mathcal{H}$
separating $X$ and line$(v,P_X(v))$ properly.
As a result, $\mathcal{H}$ must contain line$(v,P_X(v))$.
Let $\textbf{n}$ be the unit normal vector
of $\mathcal{H}$ with $\angle(\textbf{n},z-P_X(v))\geq\pi/2$ for any $z\in X$,
and $\mathcal{H}_{v}$ the tangent plane of bd$(X)$ at $P_X(v)$.
Then $\textbf{n}\in\mathcal{H}_{v}$ since $v-P_X(v)$ is a normal vector of tangent plane $\mathcal{H}_{v}$.
However, it is not possible
that there exists a curve on bd$(X)$ with tangent vector $\textbf{n}$ at $P_X(v)$,
which yields a contradiction.

We now show the conclusion by contradiction.
Suppose that there is a sequence $\{v_k\}_{k\geq0}$
with $v_k\in S\backslash X$ and $P^a_{X}(v_k)\neq P_{X}(v_k)$ such that $\lim_{k\rightarrow\infty}\mu(v_k)=0$.
Without loss of generality, we assume $\lim_{k\rightarrow\infty}v_k=:v^*\in S\backslash$int$(X)$.

We first consider the case of $v^*\in S\backslash X$.
In the case of $P^a_{X}(v^*)\neq P_{X}(v^*)$, by the continuity we have
$0=\lim_{k\rightarrow\infty}\mu(v_k)=\mu(v^*)>0$, which yields a contradiction.
In the case of $P^a_{X}(v^*)=P_{X}(v^*)$, we have $\lim_{k\rightarrow\infty}\theta(v_k)=0$,
which implies
$\lim_{k\rightarrow\infty}\vartheta(v_k)=\pi$ along with $\lim_{k\rightarrow\infty}\mu(v_k)=0$. This, however, is impossible since
the surface bd$(X)$ is regular.

We next consider the case of $v^*\in$bd$(X)$.
Let $r>0$ be a sufficiently small number such that
(\ref{co}) holds for $\theta^*$ and any $v+r\textbf{n}(v)$
with $|v-v^*|\leq r$ and $v\in$bd$(X)$, where
$\textbf{n}(v)$ is the unit normal vector of the tangent plane of bd$(X)$ at $v$.
Denote $z:=v+r\textbf{n}(v)$.
Take arbitrarily a point
$\hat y:=\hat y(z)\in$ bd$(\mathbf{C}_X(z,\theta^*))\cap\mathbf{b}(z,X)\cap$aff$\{v,z,P^a_X(z)\}$
such that $\angle(v-z,\hat y-z)=\theta^*$.
Then
\begin{align}\label{as}
\mu(z)\geq\angle(v-\hat y,z-\hat y).
\end{align}
Moreover, it is not hard to find that, for any $z_1,z_2$
such that $z_1\not\in X$, $z_2\not\in X$, $P_X(z_1)=P_X(z_2)$, $|z_2|_X>|z_1|_X$,
and with (\ref{co}) holding for both ($z_1$, $\theta^*$) and ($z_2$, $\theta^*$),
we have
\begin{align}\label{sa}
\beta(z_1)\geq\beta(z_2),
\end{align}
where $\beta(z)=\inf_{\bar y\in bd(\mathbf{C}_X(z,\theta^*))\cap\mathbf{b}(z,X)}\angle(P_X(z)-\bar y,z-\bar y)$.

From (\ref{as}) and (\ref{sa}) we conclude that for any sufficiently large
$k$,
$$
\mu(v_k)\geq\inf_{v\in bd(X),|v-v^*|\leq r}\beta(v+r\textbf{n}(v))>0,
$$
which yields a contradiction. We complete the proof.
\hfill$\square$

\vspace{5mm}
\noindent \textsc{Youcheng Lou} \\
\noindent Department of Systems Engineering and Engineering Management\\
\noindent The Chinese University of Hong Kong \\
\noindent Shatin, N.T., Hong Kong \\
\noindent \textsc{} \\
\noindent Academy of Mathematics and Systems Science\\
\noindent Chinese Academy of Sciences \\
\noindent Beijing 100190, China \\
\noindent Email: {\tt\small yclou@se.cuhk.edu.hk, louyoucheng@amss.ac.cn}

\vspace{15mm}
\noindent \textsc{Yiguang Hong and Shouyang Wang} \\
\noindent Academy of Mathematics and Systems Science\\
\noindent Chinese Academy of Sciences \\
\noindent Beijing 100190, China \\
\noindent Email: {\tt\small yghong@iss.ac.cn, sywang@amss.ac.cn} \\

\end{document}